\documentclass[journal,10pt,twocolumn]{IEEEtran}
\makeatletter
\def\ps@headings{%
\def\@oddhead{\mbox{}\scriptsize\rightmark \hfil \thepage}%
\def\@evenhead{\scriptsize\thepage \hfil \leftmark\mbox{}}%
\def\@oddfoot{}%
\def\@evenfoot{}}
\makeatother
\usepackage{stfloats}
\usepackage{amsfonts}
\usepackage{multirow}
\usepackage{diagbox}
\usepackage{longtable}
\usepackage{graphicx}
\usepackage{CJK}
\usepackage{amssymb}
\usepackage{mathrsfs}
\usepackage{times}
\usepackage{amsmath,amssymb,latexsym}
\usepackage{epsfig}
\usepackage{tabularx}
\usepackage{ifpdf}
\usepackage{cite}
\usepackage{array}
\usepackage{mdwmath}
\usepackage{CJK}
\usepackage{mdwtab}
\usepackage{eqparbox}
\usepackage{subfigure}
\usepackage{color}
\usepackage{url}
\usepackage{graphicx} 
\usepackage{epstopdf}
\usepackage{morefloats}
\usepackage{subfigure}
\usepackage[font=small]{caption}
\usepackage{mathrsfs}
\usepackage{times}
\usepackage{amsmath,amssymb,latexsym}
\usepackage{epsfig}
\usepackage{tabularx}
\usepackage{ifpdf}
\usepackage{cite}
\usepackage{array}
\usepackage{mdwmath}
\usepackage{mdwtab}
\usepackage{eqparbox}
\usepackage{subfigure}
\usepackage[justification=centering]{caption}
\usepackage{color}
\usepackage{slashbox}
\usepackage{url}
\usepackage{textcomp}
\usepackage{color}
\usepackage{subfigure}
\usepackage{graphicx}
\usepackage{cite}
\usepackage{mathrsfs}
\usepackage{times}
\usepackage{amsmath,amssymb,latexsym}
\usepackage{epsfig}
\usepackage{tabularx}
\usepackage{ifpdf}
\usepackage{cite}
\usepackage{array}
\usepackage{mdwmath}
\usepackage{mdwtab}
\usepackage{eqparbox}
\usepackage{subfigure}
\usepackage{color}
\usepackage{url}
\usepackage{amsthm}
\usepackage{morefloats}
\usepackage{float}
\usepackage{cite}
\usepackage{booktabs}
\usepackage{cases}
\usepackage{cite}
\usepackage{amsmath,amssymb,amsfonts}
\usepackage{algorithmic}
\usepackage{graphicx}
\usepackage{textcomp}
\usepackage{xcolor}
\usepackage{lettrine}
\usepackage{amsmath}
 \usepackage{indentfirst} 
 \usepackage{subfigure}
 \usepackage{amsfonts}
  \usepackage[T1]{fontenc}
  \usepackage[utf8]{inputenc}
\makeatletter
\usepackage{algorithmic}
\usepackage{algorithm}
\usepackage{verbatim}
\renewcommand{\maketag@@@}[1]{\hbox{\m@th\normalsize\normalfont#1}}%
\makeatother
\begin{document}

\newtheorem{lemma}{Lemma}
\title{Dual-Scale Channel Estimation
in Sensing-Assisted Communication Systems: Joint Time Allocation and Beamforming Design}
\author{Zhiyue Bai, Minghui Dai,~\IEEEmembership{Member,~IEEE},  Fen~Hou,~\IEEEmembership{Member,~IEEE}, Hangguan~Shan,~\IEEEmembership{Senior Member,~IEEE},\\ Lin X.~Cai,~\IEEEmembership{Fellow,~IEEE}, and Xuemin (Sherman)~Shen,~\IEEEmembership{Fellow,~IEEE}
\thanks{Part of this work was presented at the 2025 IEEE/CIC International Conference on Communications in China (ICCC)\cite{bai2025optimal}, Shanghai, China, August 2025.} 
 \thanks{Zhiyue Bai and Fen Hou are with the State Key Laboratory of Internet of Things for Smart City, the Department of Electrical and Computer Engineering, University of Macau, Macau, China (e-mail: yc27410@um.edu.mo; fenhou@um.edu.mo).}
 \thanks{Minghui Dai is with the School of Computer Science and Technology, Donghua University, Shanghai 201620, China (e-mail: minghuidai@dhu.edu.cn).}
\thanks{Hangguan Shan is with the College of Information Science and Electronic Engineering, Zhejiang University, Hangzhou 310027, China (e-mail: hshan@zju.edu.cn).}
\thanks{Lin X. Cai is with the Department of Electrical and Computer Engineering, Illinois Institute of Technology, Chicago, IL 60616 USA (e-mail: lincai@iit.edu).}
\thanks{Xuemin (Sherman) Shen is with the Department of Electrical and Computer Engineering, University of
Waterloo, Waterloo, ON N2L 3G1, Canada (Email: sshen@uwaterloo.ca).}
}

\maketitle

\begin{abstract}
In this paper, we propose a novel integrated sensing and communication (ISAC)-enabled dual-scale channel estimation framework, where large-scale channel estimation benefits from sensing, and the temporal variation of small-scale channel state information is modeled via channel aging. By characterizing the impact of angular sensing error on the communication spatial correlation matrix, we derive a closed form expression for the achievable rate under dual-scale channel estimation errors. Considering the different characteristics in time scales, we design the sensing duration for slow-varying large-scale channel and determine the update timing and frequency for fast-varying small-scale channel information within a given frame structure. We formulate an average achievable rate maximization problem under limited time resources and sensing Cramer-Rao bound (CRB) constraints, and propose a segmented golden based joint optimization algorithm to efficiently solve this nonconvex problem. Simulation results demonstrate that our proposed scheme achieves significant performance improvement compared with the benchmark schemes, which further validate that the system can leverage additional sensing capabilities to enhance communication efficiency.
\end{abstract}
  
  \begin{IEEEkeywords}
  Integrated sensing and communication, channel estimation, resource allocation, beamforming.
  \end{IEEEkeywords}

  \section{Introduction}
  \IEEEPARstart{T}{he} interaction between sensing and communication (S\&C) in widespread applications has positioned it as a key topic in 6G, sparking the rapid development of integrated sensing and communication (ISAC)\cite{wang2024generative}. On one hand, the resource competition and mutual interference arising from spectrum conflict between S\&C in 6G are increasingly significant, necessitating the coexistence or integrated architecture designs to ensure system stability\cite{zhang2021overview}. On the other hand, with ubiquitous application scenarios,  such as 
  real-time disaster monitoring and prediction, security monitoring and behavior detection in smart homes, as well as high-precision and low-latency task processing in Industrial Internet of Things\cite{shao2025intelligent}, the promising potential of ISAC continues to attract substantial research interest.  Research on technical details such as channel estimation design under specific frame structure and collaborative mechanisms between S\&C in ISAC is still ongoing and requires further advancement~\cite{zhang2021overview, liu2020radar}.\par
  ISAC provides two potential gains: {\romannumeral1}) the integration gain derived from the shared utilization of wireless resources and hardware infrastructure. {\romannumeral2}) the coordination gain derived from the mutual assistance between S\&C~\cite{liu2022integrated}.
  Owing to the significant similarities in hardware architecture and signal processing algorithms between S\&C systems, many studies focus on enhancing integration gain\cite{liu2022integrated,zhang2021overview}. Liu et al. in~\cite{8288677} demonstrated the feasibility of antennas to transmit a joint waveform for both radar and communication in  multiple-input multiple-output (MIMO) systems through antenna design. Since the deterministic sensing signal fails to carry the random communication information,  some studies focused on communication-centric design\cite{zhang2021overview,chen2023impact}, and Bai et al. in \cite{bai2023optimal} proposed a passive sensing model for vehicle to vehicle cooperation using communication pilot signals.  As the research on millimeter-wave communication advances, the high signal attenuation and narrowband characteristics substantially increase the time overhead required for beam alignment, thereby motivating the investigation of sensing-assisted communication systems\cite{wang2025generative}.  Tan et al. in \cite{tan2024beam} demonstrated  that robust location information obtained from radar can significantly reduce the overhead for beam alignment and prediction in millimeter-wave communication. Yuan et al. in~\cite{yuan2020bayesian} proposed a low-overhead predictive beamforming scheme for ISAC-enabled vehicular networks, leveraging Bayesian inference to achieve real-time motion parameter estimation with reduced signaling overhead.  Leveraging the impact of sensing angular accuracy on communication, Fan et al. in \cite{fan2022radar} proposed a radar-integrated MIMO communication scheme for multi-hop vehicle-to-vehicle networks, jointly optimizing power allocation and link selection to reduce power consumption/outage probability. Meng et al. in \cite{meng2023sensing} proposed a novel sensing-assisted communication scheme for vehicular networks by deploying an intelligent omni-surface on vehicles to enhance both S\&C performance. Shao et al. in \cite{11165349}, considered slow-timescale acquisition of user location information, which was then exploited at the fast timescale to obtain the instantaneous CSI. However, existing research lacks a quantitative analysis about how sensing accuracy impacts communication efficiency. In this work, we aim to study the relationship between sensing accuracy and communication efficiency from the information-theoretic perspective.\par 

The communication channel comprises both large-scale information and small-scale information. Thus, efficient channel estimation must ensure reliable acquisition of information on both scales. While sensing aids in acquiring large-scale parameters like angle and distance, traditional pilot-based methods remain essential for capturing  small-scale information~\cite{gao2022integrated}. Due to the influence of multipath effects and Doppler shift, the coherence time of fast-varying small-scale fading can be as short as a few milliseconds in mobile scenarios~\cite{jiang2022accurate}.  Hence, the study on dual-scale channel estimation needs to be further explored. To characterize the small-scale fading, a channel aging model based on the maximum Doppler shift is widely adopted~\cite{xia2020learning,chen2023impact, zheng2021impact}. Zheng et al. in \cite{zheng2021impact} employed the channel aging model to analyze the uplink and downlink achievable rates under imperfect time-varying channel estimation, highlighting the superiority of cell-free massive MIMO systems over small-cell systems. Nevertheless, the study of channel aging in sensing-assisted communication systems deserves further investigation.\par
Motivated by the above discussions, we aim to explore the optimal design for dual-scale channel estimation in sensing-assisted communication systems. In this paper, we address the following two key issues. Firstly, how can the impacts of sensing performance on communication  be explicitly quantified through a closed-form expression? This involves the construction of a mathematical relationship between imperfect channel estimation and sensing metric. Moreover, the trade-off between S\&C functionalities in sensing-assisted communication systems remains an area worthy of further investigation. Secondly, how can dual-scale channel estimation be effectively designed?  Inspired by the aforementioned challenging issues, we model and analyze the impacts of dual-scale channel estimation on the system performance such as the achievable rate.  Based on dual-scale channel estimation, we propose the optimal time and power scheduling strategies for the ISAC system. In summary, the main contributions of this paper are summarized as follows.
  \begin{itemize}
  \item\textit{Dual-Scale Channel Estimation for Sensing-Assisted Communication Systems:} Considering both the large-scale and the small-scale channel estimations, we investigate a sensing-assisted communication system in which sensing is used for large-scale information detection, and the minimum mean squared error (MMSE)  estimator is used for small-scale estimation. Furthermore, we quantify small-scale block fading effects through the channel aging model, and the temporal difference between the two scales motivates  our investigation of a frame-structured design.  In a scheduling cycle, we determine the sensing duration to derive the  slow-varying large-scale information, and the update timing and frequency of fast-varying small-scale estimation. Our goal is to maximize the system's average achievable rate through the optimization of the dual-scale time resources, radar transmit beamformers, and communication transmission power.
  \item\textit{Imperfect Channel Estimation Model:} Considering the impact of dual-scale channel estimation errors on communication efficiency, we explore the effects of channel aging in small-scale fading  and  sensing accuracy  in large-scale fading for channel modeling. By employing the MMSE estimator and maximum ratio transmission (MRT) beamforming, we derive the achievable rate with the consideration of dual-scale channel estimation.  Furthermore,  we analyze the collaboration of S\&C in enhancing system performance, i.e., leveraging additional sensing capabilities to improve communication efficiency. 
  \item\textit{Optimal Design for System Performance:} With the alternating optimization (AO) method, we decouple  the original problem into two subproblems: time-related optimal design and power-related optimal design. By applying the Lagrange multiplier method, we derive the closed-form solutions of small-scale update timing. We further prove that the time-related  subproblem is a segmented convex problem and employ a golden search method to determine the optimal sensing duration and small-scale update frequency.  For the power-related optimal design, we reformulate the subproblem into a convex semidefinite programming (SDP) problem through mathematical transformations. 
  \item\textit{Effectiveness of the Proposed Scheme:} Simulation results validate the superiority of our proposed scheme over four benchmark schemes, achieving performance improvements ranging from 10.7\% to 152.4\%. In mobile networks, the optimization of small-scale update frequency and sensing duration significantly enhances system performance. Moreover, our results reveal the cooperative gain  introduced by sensing-assisted communication.\par
  \end{itemize}
  \qquad The rest of this paper is organized as follows.  
  Section \ref{SystemModel} and \ref{ProblemFormulation} present the system model and the problem formulation, respectively. 
  Section \ref{ProblemDecomposition} provides the solvable subproblems and introduces the proposed efficient algorithm for the dual-scale channel estimation. Numerical results are presented in Section \ref{PerformanceEvaluation} and the paper is concluded in Section \ref{Conclusion}.\par
  \textit{Notations}: Boldface upper-case and lower-case letters denote matrix and vector, respectively. $ \mathbb{C}^{d_1\times d_2}$ stands for the set of complex $ d_1\times d_2$ matrices. $\mathbb{E}\left( \cdot \right)$, $\left( \cdot \right)^T$, $\left( \cdot \right)^*$,  $\left( \cdot \right)^H$, $\left( \cdot \right)^{-1}$  represent the expectation, transpose, conjugate,  transpose-conjugate, and inverse operation, respectively. For a matrix  $\boldsymbol{X}$, $\mathrm{tr}\left( \boldsymbol{X} \right)$, $\mathrm{rank}\left( \boldsymbol{X} \right)$, $\mathrm{Re}\left\{ \boldsymbol{X} \right\}$, and $\mathrm{Im}\left\{ \boldsymbol{X} \right\}$ stand for its trace, rank, real part, and imaginary part, while $\boldsymbol{X}\succeq 0$ indicates that $\boldsymbol{X}$ is positive semi-definite. $\mathrm{diag}\left( \boldsymbol{x} \right)$ and $\mathrm{diag}\left( \boldsymbol{X} \right)$ denote the diagonal matrixes whose main diagonal elements are the elements $\boldsymbol{x}$, and diagonal elements of  $\boldsymbol{X}$, respectively. $\lfloor \ \rfloor$ is the floor function and $\mathcal{O} \left( \cdot \right)$ is the big-O computational complexity notation.  

  \begin{figure}[htbp]
    \centerline{\includegraphics[width=8cm]{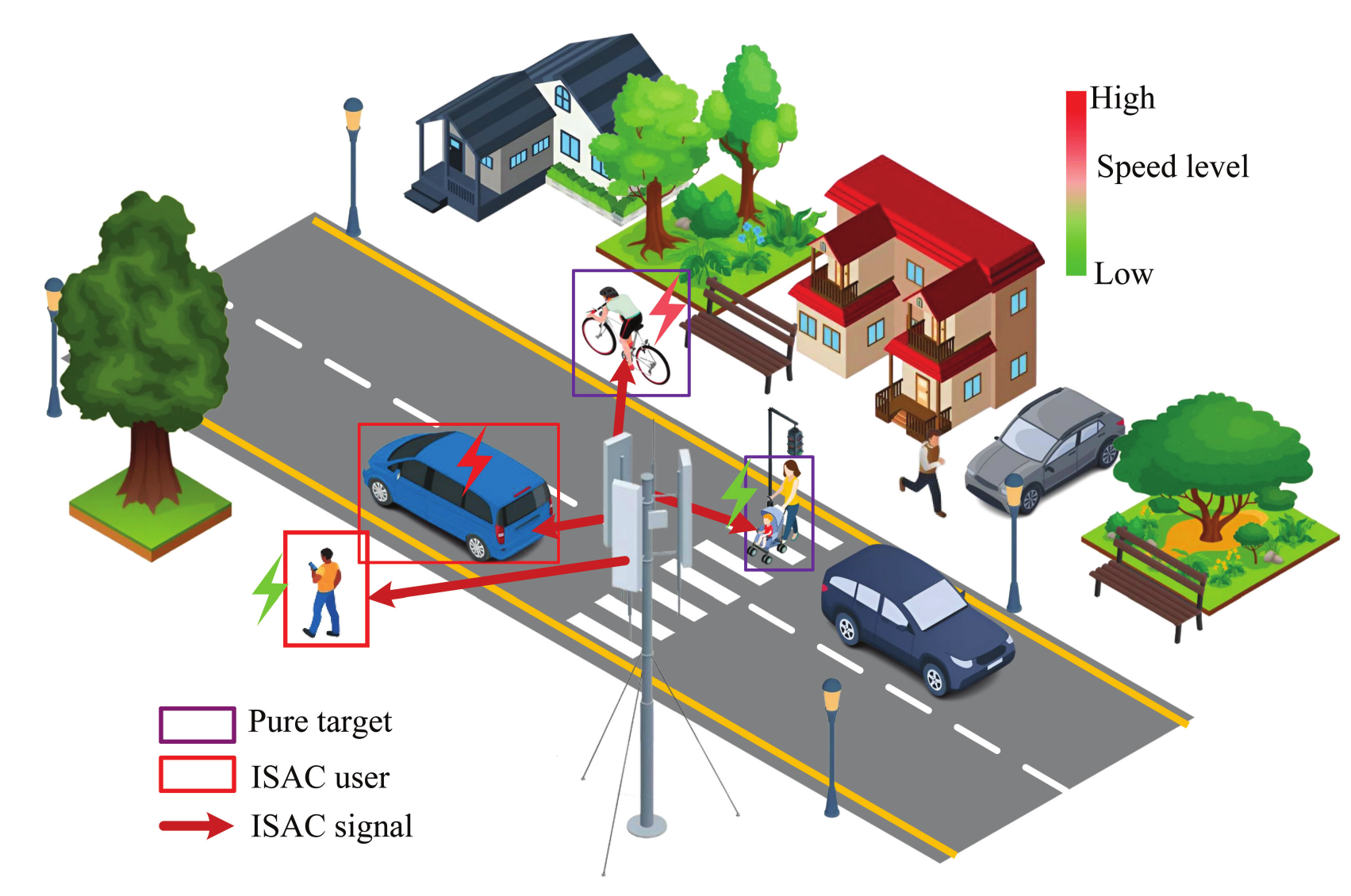}}
    \caption{A scenario of sensing-assisted communication system considering user mobility.}
    \vspace{-1.5em}
    \label{paper3}
  \end{figure}
  \section{System Model and Problem Formulation}\label{SystemModel}
  As shown in Fig. \ref{paper3}, we consider ISAC-enabled downlink transmission consisting of a dual-function base station (BS),  $L$ pure sensing targets, and $K$ ISAC users.  The BS employs a
  general uniform linear array (ULA) with $L_t$ transmit antennas for communication and tracking targets, and $L_r$ receive antennas for sensing echo signal reception. The pure sensing targets, denoted by 
  $\mathcal{L} =\left\{ 1, 2, ..., L \right\}$, could be some important sensing locations, e.g., the pedestrian crossing shown in Fig. \ref{paper3}. For the ISAC users denoted by $\mathcal{K} =\left\{ 1, 2,..., K \right\}$,  radar sensing results can be used to enhance the communication channel estimation, and then assist in the improvement of communication efficiency. \par
In this paper, we consider dual-scale channel estimation, where the estimation of the angle-dependent spatial correlation matrix is termed as large-scale channel estimation, while the estimation of the random components affected by multipath and Doppler shifts is termed as small-scale channel estimation. Fig.~\ref{frame} illustrates the frame structure with dual-scale channel estimation. At the beginning of each subframe, large-scale channel estimation will be conducted through a radar sensing process. The obtained slow-varying large-scale spatial correlation matrix is key to the efficient precoding and resource scheduling in communication systems. In addition, the small-scale Doppler shifts and the mobility of users lead to channel aging, which makes the small-scale channel information across the transmission blocks  inconsistent and time-correlated\cite{xia2020learning,zheng2021impact}. Therefore, multiple small-scale channel estimations should be arranged in the subframe.  The increase in the time duration of large-scale sensing and the frequency of small-scale updates can improve the channel estimation accuracy. However, it also sacrifices the data transmission time. Therefore, the optimal design of these variables is critical and challenging in an ISAC system. The frame structure is elaborated in the following subsection.\par
  \begin{figure}[htbp]
    \hspace{-1em}
    \centerline{\includegraphics[width=7.8cm]{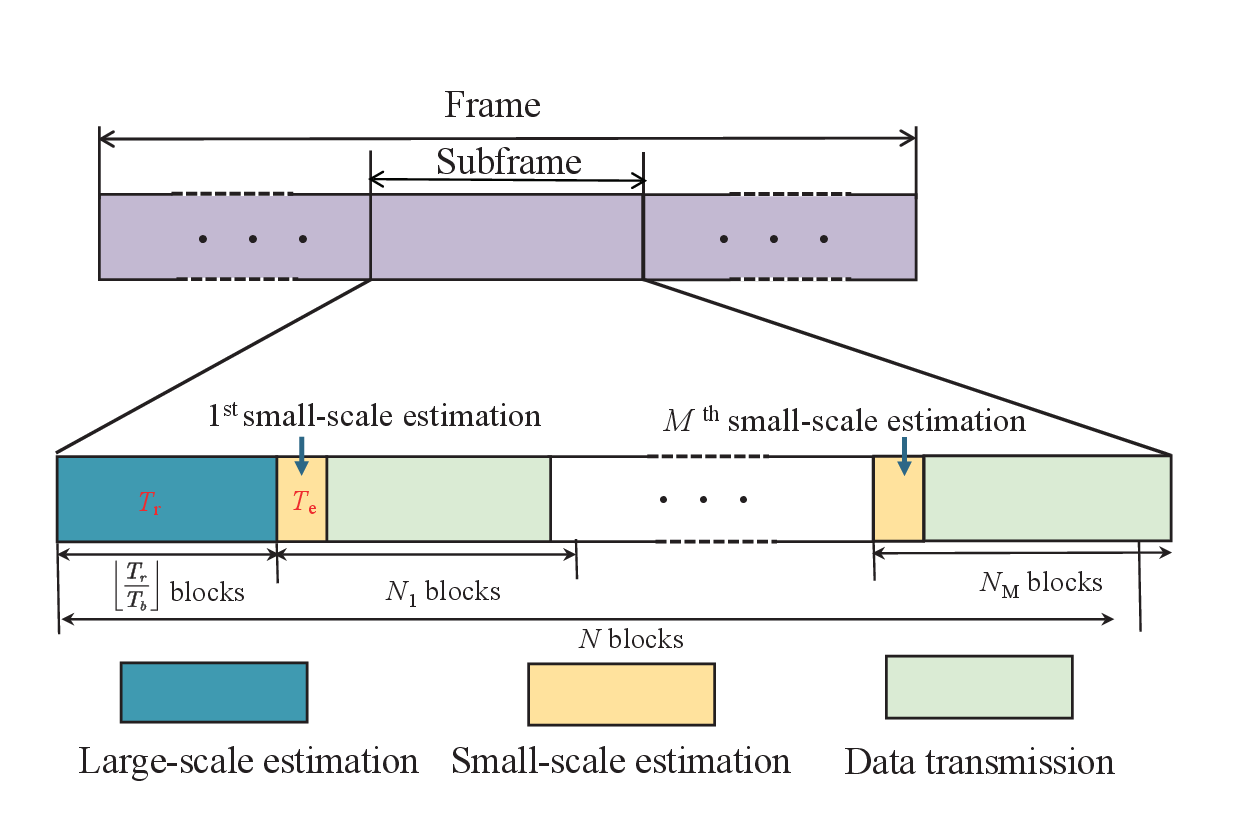}}
    \caption{Illustration of the frame structure with dual-scale channel estimation.}
    \label{frame}
    \vspace{-1em}
  \end{figure}
  \subsection{Frame Structure}
    As shown in Fig. \ref{frame}, each frame is divided into multiple subframes, and each subframe consists of $N$ blocks. The beginning of each subframe is subjected to a large-scale detection periodically, which is consistent with the design of the physical layer  channel state information reference signal (CSI-RS). The acquisition of large-scale slow-varying channel information is generally achieved by designing the ``CSI-ResourcePeriodicityAndOffset'' parameter to be periodic between 4-640 timeslots\cite{dahlman20205g}. The time duration of radar-derived large-scale detection is denoted as $T_r$. In addition, several small-scale estimations, denoted as $M$,  should be arranged within each subframe. The time duration of each small-scale estimation is typically fixed and denoted as $T_e=Q_e T_s$, where $Q_e$ is the number of symbols in each small-scale estimation and $T_s$ is the time duration of a symbol. $Q_e$ is also termed as the small-scale channel estimation overhead per update.  This is consistent with the design of the demodulation reference signal (DM-RS) reference signals\cite{38.221}. \par
    The time duration of a block is denoted as  $T_b=Q_bT_s$,  where $Q_b$ is the number of symbols in the block and $T_s$ is the time duration of each symbol. 
    Within each subframe of $N$ blocks,  $\lfloor \frac{T_r}{T_b} \rfloor$ blocks are allocated for radar sensing to achieve large-scale channel detection. It is noted that the sensing duration may span multiple transmission blocks. The remaining  $N_t=N-\lfloor \frac{T_r}{T_b} \rfloor$ blocks are arranged with $M$ small-scale channel estimations.  The time duration between $m$-th small-scale estimation and ($m+1$)-th small-scale estimation is called as the time duration of $m$-th small-scale update, denoted as $N_m$ blocks, $m\in \left\{ 1,2,...., M \right\}$.
    To improve the achievable rate, this system entails a compromise on several time variables, i.e., the duration of large-scale detection $T_r$, the number of small-scale estimations $M$, and the number of blocks for each small-scale update $N_m$. 
    \subsection{Channel Model}
    We consider that the channel is quasi-static in each block but varies across different blocks. From a practical standpoint, we adopt the classical ``one-ring'' channel model, where the spatial correlation matrix captures the large-scale statistics and the Rayleigh fading denotes the random scatterers around the user at small-scale. The time-varying channel for user $k$ at block $n$ of $m$-th update can be expressed as \cite{ma2018sparse, xia2020learning}
    \begin{small}
      \begin{equation}
 \begin{split}
    \label{62614}
&\boldsymbol{h}_{k,n,m}=\int_{-\infty}^{+\infty}{\int_{\theta _k-\Delta \theta}^{\theta _k+\Delta \theta}{\boldsymbol{a}\left( \theta \right)}e^{\left( j2\pi f_dT_bn \right)}r_k\left( \theta ,f_d \right) d\theta df_d}, \\& \qquad \qquad\qquad\qquad\qquad\qquad\qquad\qquad\qquad\quad \qquad\!\!\!\forall k,n,m,
\end{split}
\end{equation}
\end{small}where $\Delta \theta$ is the single-side angular spread (AS), and $r_k\left( \theta,f_d \right)$ denotes the joint angle-Doppler channel gain function on the direction of arrival (DOA) $\theta$ and the Doppler frequency $f_d$. 
Leveraging the sizable altitude of the BS and the narrow-beam nature of millimeter-wave, the AS will be limited in a narrow region, compared to the conventional model. Then, to derive the spatial correlation matrix of $\boldsymbol{h}_{k,n,m}$, it is general to assume that the joint angle-Doppler channel gain $r_k\left( \theta,f_d \right)$ in (\ref{62614}) is uncorrelated  for different ($\theta$, $f_d$) 
\begin{equation}
      \begin{split} 
        \label{62615}
     \mathbb{E}  &\left[ r_k\left( \theta ,f_d \right) r_{k}^{*}\left( \theta ^{\prime},f_{d}^{\prime} \right) \right]\\
      &= \beta _k S_{k}^{A}\left( \theta \right) S_{k}^{D}\left( f_d \right) \delta \left( \theta -\theta ^{\prime} \right) \delta \left( f_d-f_{d}^{\prime} \right), \forall k,
      \end{split} 
\end{equation}
where $\beta _k$,  $S_{k}^{A}\left( \theta \right)$, and $S_{k}^{D}\left( f_d \right)$ are the average channel gain, power-angle-spectrum, and power-Doppler-spectrum, respectively\cite{clerckx2013mimo}. Then, by plugging  (\ref{62615}) into (\ref{62614}),  the spatial correlation matrix of $\boldsymbol{h}_{k,n,m}$ for any $k,n,m$  is given by 
  \begin{small}
     \begin{equation}
 \begin{split}
  \label{11303}
 \mathbb{E}&\left[ \boldsymbol{h}_{k,n,m}\boldsymbol{h}_{k,n,m}^{H} \right] \\& =\beta _k\int_{\theta _k-\Delta \theta}^{\theta _k+\Delta \theta}{S_{k}^{A}\left( \theta \right) \boldsymbol{a}}\left( \theta \right) \boldsymbol{a}^H\left( \theta \right) d\theta\triangleq \beta _k\boldsymbol{R}_{k}^{S},\forall k,n,m,
\end{split}
\end{equation} 
  \end{small}where $\boldsymbol{R}_{k}^{S}\in \mathbb{C}^{L_t\times L_t}$ is the large-scale spatial correlation matrix\footnotemark{}.
Thus, we have $\boldsymbol{h}_{k,n,m}\sim \mathcal{C}\mathcal{N}\left( 0,\beta _k\boldsymbol{R}_{k}^{S} \right)$.  It can be observed that $\boldsymbol{h}_{k,n,m}$ depends only on the indices $k$ and $n$, where $n$ is related to the Doppler frequency (small-scale). The large-scale statistics in  (\ref{11303}) depend only on the user index $k$. Therefore, for any update $m$,  $\boldsymbol{h}_{k,n,m}$ has identical statistical properties. Thus, for notational convenience, we omit the subscript $m$.  The following description is valid for any update duration $m\in M$, unless particularly noted.\par
\footnotetext{Given the temporal scope of resource scheduling spanning tens of milliseconds, we consider that the large-scale characteristics of the communication channel remain constant\cite{huang2021mimo,biguesh2006training}. Thus, the large-scale spatial matrix $\boldsymbol{R}_{k}^{S}$ is independent of the time index.}
Let $\boldsymbol{h}_{k,1}$ and $\boldsymbol{h}_{k,n}$ be the physical ground-truth channel  for the $1^{st}$ and $n^{th}$ clock at the duration of $m^{th}$ small-scale channel update, respectively. From the statistical point of view, they follow the same distribution $\boldsymbol{h}_{k,n}\sim \mathcal{C} \mathcal{N} \left( 0,\beta _k\boldsymbol{R}_{k}^{S} \right)$.  However, in the specific realizations, the small-scale components of  $\boldsymbol{h}_{k,1}$ and $\boldsymbol{h}_{k,n}$ vary from block to block due to block fading. After the small-scale estimation at the beginning of $m$-th small-scale update, we obtain the small-scale channel information at the nearest time instant, i.e., $\boldsymbol{h}_{k,1}$ is known. Then, at the following time duration,  the small-scale component varies due to block fading until the update of next small-scale channel estimation. Therefore, we can only use $\boldsymbol{h}_{k,1}$ to estimate $\boldsymbol{h}_{k,n}$ through the channel aging model, which is given by (omitting the index $m$) 
    \begin{equation}
      \label{7318}
      \boldsymbol{h}_{k,n}=\rho _{k,n}\boldsymbol{h}_{k,1}+\boldsymbol{e}_{k, n}, \forall k,n,
    \end{equation}
  where $\rho _{k,n}\boldsymbol{h}_{k,1}$ represents the estimate of $\boldsymbol{h}_{k,n}$ based on $\boldsymbol{h}_{k,1}$.  The second term  $\boldsymbol{e}_{n, k}$ represents the estimation error of the round-truth value $\boldsymbol{h}_{k,n}$.  $\rho _{k,n}$ denotes the temporal correlation factor, which decreases as the time interval increases. The value of  $\rho _{k,n}$ can be obtained via the temporal-spatial autocorrelation matrix~\cite{ma2018sparse}
   \begin{small}
    \begin{equation}
    \begin{split}
        \label{1123}
        \mathbb{E}\left[ \boldsymbol{h}_{k,1}\boldsymbol{h}_{k,1+n}^{H} \right] =\underset{\rho _{k,n}^{2}}{\underbrace{\int_{-\infty}^{+\infty}{S_{k}^{D}\left( f_d \right) e^{-2j\pi f_dT_bn}}df_d}}\times \beta _k\boldsymbol{R}_{k}^{S}, \forall k,n.
    \end{split}
    \end{equation}
  \end{small}Therefore, the estimation of $\boldsymbol{h}_{k,n}$ is {\small$\rho _{k,n}\boldsymbol{h}_{k,1}\sim \mathcal{C}\mathcal{N}\left( 0,\rho _{k,n}^{2}\beta _k\boldsymbol{R}_{k}^{S} \right)$,} and {\small$\boldsymbol{e}_{k, n} \sim \mathcal{C} \mathcal{N} \left( 0,\beta _k \left( 1-\rho _{k,n}^{2} \right) \boldsymbol{R}_{k}^{S} \right)$} is the independent residual error.
Based on the Clarke-Jakes model\cite{chen2023impact}, the closed-form solution of $\rho _{k,n}$ is given by
  \begin{equation}
      \label{22416}
      \rho _{k,n} =J_0\left( 2\pi f_{d,k}^{\max}T_b\Delta n \right), \forall k,n,
    \end{equation}
where $J_0\left( \cdot \right)$ is the zeroth-order Bessel function of the first kind, and $f_{d,k}^{\max}$ is the maximum Doppler frequency of user $k$.\par
\subsection{Dual-Scale Channel Estimation}
In this subsection, we analyze and derive both the large-scale and small-scale components of the communication channel through dual-scale channel estimation. Since the random scattering around the user also exhibits spatial structure, the small-scale estimation must be built upon the large-scale spatial correlation matrix, and we therefore first estimate the large-scale component $\beta _k\boldsymbol{R}_{k}^{S}$.
\subsubsection{Large-Scale Detection Model}
In an ISAC system, the large-scale channel estimation conducted at the BS depends on the radar sensing accuracy\footnotemark{}. The sensing seudorandom sequence $\mathbf{s}_r\left( t \right) =\left[ s_{r,1}\left( t \right) , ... , s_{r,K}\left( t \right) \right] ^T\in \mathbb{C}^{K\times 1}$ is transmitted as $\boldsymbol{\tilde{s}}_r\left( t \right) =\boldsymbol{W}_{r}\boldsymbol{s}_r\left( t \right) \in \mathbb{C}^{L_t\times 1}$ for ISAC users' tracking at  time $t$, where $\boldsymbol{W}_{r}\in \mathbb{C}^{L_t\times K}$ is the transmit beamforming matrix for target detection. We consider that the ISAC signals $\left\{ s_{r,k}\left( t \right) \right\}$ for different targets are orthogonal, i.e., $ s_{r,k}\left( t \right) s_{r,j}\left( t \right) =0$, where $k \ne j$, and  $ s_{r,k}\left( t \right) s_{r,k}^*\left( t \right) =1$,  $k,j\in \mathcal{K}$.  Then, the reflected echoes at BS are given by
\begin{equation}
 \begin{split}
  \boldsymbol{y}_r\left( t \right)=&\sum_{k=1}^K{\sqrt{L_rL_t}}\alpha _{k}e^{j2\pi \nu _{k}t}\boldsymbol{b}\left( \theta _{k} \right) \boldsymbol{a}^T\left( \theta _{k} \right) \\
  &\times \boldsymbol{w}_{r,k}s_{r,k}\left( t-\tau _{k} \right)+\boldsymbol{z}_r\left( t \right), 
 \end{split}
\end{equation}
\footnotetext{In this work, we assume that preliminary sensing has already provided range and angle information, either via beam prediction and tracking or from a detection stage that optimizes the SNR \cite{liu2020radar,wu2025intelligent}.  We focus on quantifying the impact of sensing parameter estimation errors on communication efficiency in sensing-assisted communication systems.}
where $\sqrt{L_rL_t}$ is the array gain factor, $\alpha _{k}$ is the reflection coefficient determined by the round-trip radar equation. $\nu _{k}$ and $\tau _{k}$ are the Doppler frequency and time delay, respectively.  $\boldsymbol{z}_r\left( t \right) \sim \mathcal{C} \mathcal{N} \left( 0,\sigma _{r}^{2}\mathbf{I}_{L_r} \right)$ stands for the complex additive
white Gaussian noise at the sensing receiver. Assuming the half-wavelength antenna spacing at the BS, the transmit and receive steering vectors, i.e., $\boldsymbol{a}\left( \theta  \right)$ and $\boldsymbol{b}\left( \theta \right)$, are given by
\begin{subequations}
  \begin{flalign}
   \! &  \boldsymbol{a}\left( \theta \right) =\frac{1}{\sqrt{L_t}}\left[ 1,e^{-j\pi \sin \theta}, ..., e^{-j\pi \left( L_t-1 \right) \sin \theta} \right] ^T, \\
   &  \boldsymbol{b}\left( \theta \right) =\frac{1}{\sqrt{L_r}}\left[ 1,e^{-j\pi \sin \theta}, ..., e^{-j\pi \left( L_r-1 \right) \sin \theta} \right] ^T.
   \end{flalign}
\end{subequations}
Since the interference between targets is negligible in the large-scale MIMO regime and the targets can be distinguished in the delay-Doppler domain \cite{ngo2015massive
}, the BS is capable of processing each echo signal individually, as $\boldsymbol{y}_{r,k}\left( t \right)$.
 Then, the delay $\tau _{k}$ and the Doppler frequency $\nu _{k}$ can be estimated by matched-filtering according to\cite{liu2020radar}
 \begin{equation}
  \begin{split}
 \left\{ \hat{\tau}_{k},\hat{\nu}_{k} \right\} \! =\!\underset{\tau_k , \nu_k}{\mathrm{arg} \! \max}\left| \int_0^{T_r}\!\!\!{\boldsymbol{y}_{r,k}\left( t \right) s_{r,k}^{*}\left( t\!-\!\tau \right)}e^{-j2\pi \nu t}dt \right|^2,\forall k,
  \end{split}
 \end{equation}
 where $T_r$ is the sensing duration. Then, the output of target $k$ after the compensation of the matched filter across time duration $T_r$ can be expressed as
 \begin{equation}
  \begin{split}
    \label{6134}
    \tilde{\boldsymbol{y}}_{r,k}= T_r\sqrt{GL_rL_t}\alpha _{k}\boldsymbol{b}\left( \theta _{k} \right) \boldsymbol{a}^T\left( \theta _{k} \right) \boldsymbol{w}_{r,k}+\tilde{\boldsymbol{z}}_{r,k},\forall k,
  \end{split}
 \end{equation}
 where $G$ is the matched-filtering gain, which is equal to the number of symbols used for matched-filtering, $\boldsymbol{w}_{r,k}$ is the $k$-th column of $\boldsymbol{W}_{r}$, and $\tilde{\boldsymbol{z}}_{r,k}\sim \mathcal{C} \mathcal{N} \left( 0,T_r\sigma _{r}^{2} \right) $ denotes the time-accumulative measurement noise. For notational convenience, we define $\dot{\alpha}_k\triangleq\sqrt{GL_rL_t}\alpha _k$, $\boldsymbol{A}\left( \theta _k \right)\triangleq\boldsymbol{b}\left( \theta _{k} \right) \boldsymbol{a}_k^T\left( \theta _{k}\right)$. In the sequel, we drop $\theta _k$ of $\boldsymbol{A}\left( \theta _k \right)$ for easier explanation, i.e., $\boldsymbol{A}_k$.\par
\vspace{0.5em}
Then, the angle $\theta _{k}$ can be measured by the maximum likelihood estimation (MLE) or super-resolution algorithms like multiple signal classification (MUSIC) methods with the estimation error $z_{\theta,k}\sim \mathcal{C}\mathcal{N}\left( 0,\sigma _{\theta,k}^2 \right)$, where $\sigma _{\theta,k}^2$ relies on the sensing quality of the target\cite{liu2022survey,kay1993fundamentals}. 
The mean square error (MSE) serves as a measure of noise variance, e.g., for angle estimation\footnotemark{}, we have
\begin{equation}
  \begin{split} 
    \label{726}
\mathrm{MSE}\left( \theta_{k} \right)=\mathbb{E} \left[ \left( \theta _{k}-\hat{\theta}_{k} \right) \left( \theta _{k}-\hat{\theta}_{k} \right) ^H \right]=\sigma _{\theta,k}^2,\forall k.
\end{split}
\end{equation}
\footnotetext{Our model focuses on the impact of angular deviation on sensing, similar to \cite{meng2023sensing,fan2022radar}. This is due to the consistent influence of sensing accuracy of speed, angle, and distance on resource scheduling.}
Cramér-Rao bound (CRB) is widely adopted as the lower bound of MSE to evaluate the sensing accuracy. Assuming that the unbiased measurements can be realized, the angular CRB of the signal model in (\ref{6134}) can be expressed as
 \begin{equation}
  \begin{split} 
    \label{728}
    \mathrm{MSE}\left( \theta _{k} \right) \geqslant \mathrm{CRB}\left( \theta _{k} \right) =\left[ \boldsymbol{F}_{k}^{-1} \right] _{1,1},\forall k.
\end{split}
\end{equation}
Handling the unknown parameter set as {\small $\boldsymbol{\xi }_k\triangleq\left[ \theta _{k},\tilde{\boldsymbol{\alpha}}_k
^{T} \right] ^T\in \mathbb{R}^{3\times 1}$}  with {\small$\tilde{\boldsymbol{\alpha}}_k=\sqrt{GL_rL_t}\left[ \text{Re}\left\{ \alpha _{k} \right\} , \text{Im}\left\{ \alpha _{k} \right\} \right] ^T$}, each element of the Fisher information matrix (FIM) $\boldsymbol{F}_{k}$ for estimating $\boldsymbol{\xi }_k$ is given by
\begin{equation}
  \begin{split} 
    \label{6159}
\left[ \boldsymbol{F}_k\left( \boldsymbol{\xi } \right) \right] _{i,j}=\frac{2}{T_r\sigma _{r}^{2}}\mathrm{Re}\left\{ \frac{\partial \tilde{\boldsymbol{y}}_{r,k}^{H}}{\partial \boldsymbol{\xi }_i}\frac{\partial \tilde{\boldsymbol{y}}_{r,k}}{\partial \boldsymbol{\xi }_i} \right\} , i,j=1,2,3,\forall k. 
\end{split}
\end{equation}
\vspace{0.5em}
The FIM is then represented as
\begin{equation}
  \begin{split} 
\boldsymbol{F}_k=\left[ \begin{matrix}
	\boldsymbol{F}_{\theta \theta ,k}&		\boldsymbol{F}_{\theta \tilde{\boldsymbol{\alpha}}
  ,k}\\
	\boldsymbol{F}_{\theta \tilde{\boldsymbol{\alpha}}
  ,k}^{T}&		\boldsymbol{F}_{\tilde{\boldsymbol{\alpha}}
  \tilde{\boldsymbol{\alpha}}
  ,k}
\end{matrix} \right],\forall k,
\end{split}
\end{equation}
where  $\boldsymbol{F}_{\theta \theta ,k}$, $\boldsymbol{F}_{\theta \tilde{\boldsymbol{\alpha}}
,k}$, and $\boldsymbol{F}_{\tilde{\boldsymbol{\alpha}}
\tilde{\boldsymbol{\alpha}}
,k}$ are showed in Appendix A based on~\cite{song2023intelligent}, and the specific angular CRB of target $k$ is also  derived as
\begin{small}
\begin{equation}
  \begin{split}
\text{CRB}\!\left( \theta _k \right) \!=\!\frac{\sigma _{r}^{2}}{2T_r\left| \dot{\alpha}_k \right|^2\!\!\left( \!\text{tr}\!\left(\! \boldsymbol{\dot{A}}_k \boldsymbol{W}_{r,k}\boldsymbol{\dot{A}}_k^H\right)\! \!-\!\frac{\left| \text{tr}\left( \boldsymbol{A}_k \boldsymbol{W}_{r,k}\boldsymbol{\dot{A}}_k^H \right) \right|^2}{\text{tr}\left( \boldsymbol{A}_k \boldsymbol{W}_{r,k}\boldsymbol{A}_k^H \right)}\!\right)},\forall k.
\end{split}
\end{equation}
\end{small}Here, we define $\boldsymbol{W}_{r,k}\triangleq   \boldsymbol{w}_{r,k}\boldsymbol{w}_{r,k}^{\text{H}}$, with $\text{rank}\left( \boldsymbol{W}_{r,k} \right) =1$ and $\boldsymbol{W}_{r,k}\succeq 0$. With $\boldsymbol{D}_t=\mathrm{diag}\left( 0,1,2,\dots ,L_t-1 \right)$ and $\boldsymbol{D}_r=\mathrm{diag}\left( 0,1,2,\dots ,L_r-1 \right)$, the partial derivative form $\boldsymbol{\dot{A}}_k = \frac{\partial \boldsymbol{A}_k}{ \partial \theta _k}$ is given by 
\begin{small}
\begin{equation}
  \begin{split}
\boldsymbol{\dot{A}}_k \!=\! -j\pi \cos \left( \theta _k \right)\! \left( \boldsymbol{D}_r\boldsymbol{b}\left( \theta _k \right) \boldsymbol{a}_k^T\!\left( \theta _k \right) +\boldsymbol{b}\left( \theta _k \right) \boldsymbol{a}_k^T\!\left( \theta _k \right) \boldsymbol{D}_t \right),\forall k.
\end{split}
\end{equation}
\end{small}Our analysis of the radar measurement model is based on $K$ ISAC users, which is equally applicable to pure sensing targets. We can similarly define $\dot{\alpha}_l$, $\boldsymbol{A}_l$,  $\boldsymbol{\dot{A}}_l$, and $\boldsymbol{W}_{r,l}$ with $\text{rank}\left( \boldsymbol{W}_{r,l} \right) =1$ and $\boldsymbol{W}_{r,l}\succeq 0$.\par
Based on the radar measurement model in (\ref{726})-(\ref{728}), we can define\cite{kay1993fundamentals}
\begin{equation}
  \hat{\theta}_k=\theta _k+\dot{z}_{\theta,k},\forall k,
\end{equation}
where $\dot{z}_{\theta,k}\sim \mathcal{C} \mathcal{N} \left( 0,\mathrm{CRB}\left( \theta _k \right) \right)$. Since the communication channel is a function of $\theta _k$ from (\ref{62614}), i.e., $\boldsymbol{h}_{k,n}=\boldsymbol{f}\left( \theta _k \right)$, the first-order taylor expansion-based delta method can be deployed as\cite{benaroya2005probability}
\begin{subequations}
  \begin{flalign}
 \mathbb{E} \left[ \boldsymbol{f}\left( \hat{\theta}_k \right) \right] \approx  & \ \boldsymbol{f}\left( \theta _k \right),\forall k,  \label{82920a}
    \\
    \mathbb{V} \left[ \boldsymbol{f}\left( \hat{\theta}_k \right) \right] \approx & \boldsymbol{f}\prime\left( \theta _k \right) \boldsymbol{f}\prime\left( \theta _k \right) ^H\mathrm{CRB}\left( \theta _k \right)\triangleq \boldsymbol{R}_{r,k},\forall k. \label{82920b}
   \end{flalign}
\end{subequations}
With (\ref{82920a}) and (\ref{82920b}), the covariance matrix of estimation error for the large-scale statistics of communication channel is $\boldsymbol{R}_{r,k}$.
Accordingly, we have\cite{li2017channel,bjornson2017massive}
\begin{equation}
  \label{71721}
  \boldsymbol{h}_{k,1}=\hat{\boldsymbol{h}}_{k,1}+\boldsymbol{e}_{r,k},\forall k,
\end{equation}
where $\hat{\boldsymbol{h}}_{k,1}$ is the sensing-estimated channel and $\boldsymbol{e}_{r,k}$ is the large-scale channel estimation error with zero mean and covariance matrix $\boldsymbol{R}_{r,k}$. Therefore, after radar measurement, the  correlation matrix  of $\hat{\boldsymbol{h}}_{k,n}$ is given by
\begin{equation}
\begin{split}
  \hat{\boldsymbol{R}}_k&=\mathbb{E} \left[ \hat{\boldsymbol{h}}_{k,n}\hat{\boldsymbol{h}}_{k,n}^{H} \right]\!\! =\mathbb{E} \left[ \boldsymbol{h}_{k,n}\boldsymbol{h}_{k,n}^{H} \right]\!\! -\mathbb{E} \left[ \boldsymbol{e}_{r,k}\boldsymbol{e}_{r,k}^{H} \right]\\
  &=\beta _k \boldsymbol{R}_{k}^{S}-\boldsymbol{R}_{r,k},\forall k.
\end{split}
\end{equation}
It can be observed from (\ref{71721}) that $\boldsymbol{e}_{r,k}$ can not be used for small-scale channel estimation, and the large-scale component of $\boldsymbol{h}_{k,1}$ is not perfectly known. Therefore, without perfect knowledge of the large-scale channel information, the subsequent small-scale channel estimation should be based on  $\hat{\boldsymbol{R}}_k$ rather than $\beta _k\boldsymbol{R}_{k}^{S}$  (i.e.,  based on the estimation $\hat{\boldsymbol{h}}_{k,1}$ rather than the ground truth value ${\boldsymbol{h}}_{k,1}$).
\vspace{0.5em}
\par
\subsubsection{Small-Scale Estimation Model}
With the orthogonal pilot sequences of length $T_e$, the observation vector for user $k$ after signal pre-processing with the imperfect large-scale information is given by  \cite{bacci2024mmse}
\begin{equation}
  \mathbf{y}_{e,k}=T_e\sqrt{P_e}\boldsymbol{\hat{h}}_{k,1}+\boldsymbol{z}_e,\forall k,
\end{equation}
where $P_e$ is the transmit power of pilot sequence, and $\boldsymbol{z}_e\sim \mathcal{C} \mathcal{N}  \left( 0,T_e\sigma _{e}^{2} \right)$ is the time-accumulated thermal noise.  Therefore, the energy ratio between the training sequence and noise can be defined as $\gamma _e=T_eP_e/\sigma _{e}^{2}$. \par
\vspace{0.5em}
Given the previously estimated large-scale statistics, i.e., $\hat{\boldsymbol{R}}_k$, the  well-investigated MMSE-based small-scale channel estimator yields\cite{biguesh2006training}
 \begin{equation}
  \begin{split}
  \label{71725}
  &\boldsymbol{\hat{h}}_{k,1}=\boldsymbol{\tilde{h}}_{k,1}+\boldsymbol{e}_{c,k},\forall k,\\
  &\boldsymbol{\tilde{h}}_{k,1}\triangleq 
\frac{1}{T_e\sqrt{P_e}}\boldsymbol{\hat{R}}_k\left( \boldsymbol{\hat{R}}_k+\frac{1}{\gamma _e}\mathbf{I}_{\text{L}_{\text{t}}} \right) ^{-1}\mathbf{y}_{e,k},\forall k,
\end{split}
\end{equation}
where the  error covariance matrix of $\boldsymbol{e}_{c,k}$ is $\boldsymbol{C}_{e,k}\triangleq \frac{\boldsymbol{\hat{R}}_k}{\gamma _e}\left( \boldsymbol{\hat{R}}_k+\frac{1}{\gamma _e}\mathbf{I}_{\text{L}_{\text{t}}} \right) ^{-1}$. By plugging (\ref{71725}) into (\ref{71721}), the  ground-truth channel $\boldsymbol{h}_{k,1}$ can be represented as
\begin{equation}
  \boldsymbol{h}_{k,1}=\tilde{\boldsymbol{h}}_{k,1}+\boldsymbol{e}_{c,k}+\boldsymbol{e}_{r,k},\forall k.
\end{equation}
Combined with the channel aging model in (\ref{7318}), we can derive 
\begin{equation}
  \boldsymbol{h}_{k,n}=\rho _{k,n}\tilde{\boldsymbol{h}}_{k,1}+\bar{\boldsymbol{e}}_{k, n},\forall k,n,\label{9628}
\end{equation}
where $\bar{\boldsymbol{e}}_{k, n}\triangleq \rho _{k,n}\left( \boldsymbol{e}_{c,k}+\boldsymbol{e}_{r,k} \right) +\boldsymbol{e}_{k, n}$ is the  total estimation error.\par
\subsection{Communication Rate in Dual-Scale Channel Estimation}
  With the transmitted communication signal $s_{c,k,n}\left( t \right)$ at block $n$, the received signals of user $k$ is given by
  \begin{equation}
    \begin{split}
       \label{112824}
      y_{c,k,n}\left( t \right) =&\boldsymbol{h}_{k,n}^{H}\sum_{i=1}^K{\sqrt{p_{i}}}\mathbf{f}_{i,n}\,\,s_{c,k,n}\left( t \right) +z_{k,n}\left( t \right) \\
      =&\sqrt{p_{k}} \mathbb{E} \left[ \boldsymbol{h}_{k,n}^{H}\mathbf{f}_{k,n} \right] s_{c,k,n}\left( t \right) +\xi _{k,n}\left( t \right) \\
      &+\zeta _{k,n}\left( t \right) +z_{k,n}\left( t \right), \forall k,n,
    \end{split}
\end{equation}
where $z_{k,n}\left( t \right)$ is the Gaussian noise with mean zero and variance $\sigma _{c}^{2}$. To obtain an ergodic closed-form expression spectral efficiency~\cite{chen2023impact,bjornson2017massive}, we take the expectation of the desired-signal-related terms in the second and third lines of (\ref{112824}).  $\xi _{k,n}\left( t \right)$ and $\zeta _{k,n}\left( t \right)$ are the interference caused by beamforming gain uncertainty and other users, respectively, which are given by
\begin{equation}
  \begin{split}
    &\xi _{k,n}\left( t \right) =\sqrt{p_{i}}\left\{ \boldsymbol{h}_{k,n}^{H}\mathbf{f}_{k,n}-\mathbb{E} \left[ \boldsymbol{h}_{k,n}^{H}\mathbf{f}_{k,n} \right] \right\} s_{c,k,n}\left( t \right),
    \\
    &\zeta _{k,n}\! \left( t \right) =\boldsymbol{h}_{k,n}^{H}\sum_{i\ne k}^K{\sqrt{p_{i}}}\mathbf{f}_{i,n}s_{c,i,n}\left( t \right),\forall k,n.
  \end{split}
\end{equation}
Accordingly, the ergodic signal-to-interference-plus-noise
ratio (SINR) of ISAC user $k$ is given by\par
\begin{equation}
  \begin{split}
  \gamma _{k,n}=\frac{p_{k}\left| \mathbb{E} \left[ \boldsymbol{h}_{k,n}^{H}\mathbf{f}_{k,n} \right] \right|^{2}}{\mathbb{E} \left[ \left| \xi _{k,n}\left( t \right) \right|^2 \right] +\mathbb{E} \left[ \left| \zeta_{k,n}\left( t \right) \right|^2 \right] +\sigma _{c}^{2}},\forall k,n. \label{82632}
  \end{split}
  \end{equation}
To maximize the spectral efficiency, the low-complexity MRT beamforming method can be adopted as
  \begin{equation}
    \begin{split}
  \mathbf{f}_{k,n}=\frac{\tilde{\boldsymbol{h}}_{k,1}}{\sqrt{\mathbb{E} \left[ \left\| \tilde{\boldsymbol{h}}_{k,1} \right\| ^2 \right]}}=\frac{1}{\sqrt{\mathrm{tr}\left( \boldsymbol{C}_{h,k} \right)}}\tilde{\boldsymbol{h}}_{k,1}, \forall k,n,\label{9633}
  \end{split}
  \end{equation}
where $\boldsymbol{C}_{h,k}=\hat{\boldsymbol{R}}_k\left( \hat{\boldsymbol{R}}_k+\frac{1}{\gamma _e}\mathbf{I}_{{L}_{t}} \right) ^{-1}\hat{\boldsymbol{R}}_k$ is the covariance matrix of $\boldsymbol{\tilde{h}}_{k,1}$.\par
\emph{Proposition 1}: 
With the MRT transmit beamforming method in (\ref{9633}) and the predicted channel information in (\ref{9628}),  SINR $\gamma _{k,n}$ in  $n$-th block is given by
\begin{equation}
 \begin{split}
  \label{11221}
  \gamma _{k,n}\!=\!\frac{p_k\rho _{k,n}^{2}\mathrm{tr}\left( \boldsymbol{C}_{h,k} \right)}{\sum_{i=1}^K\frac{\beta _kp_i\mathrm{tr}\left( \boldsymbol{R}_{k}^{S}\boldsymbol{C}_{h,i} \right)}{\mathrm{tr}\left( \boldsymbol{C}_{h,i} \right)}\!+\!\sigma _{c}^{2}},\forall k,n,
\end{split}
\end{equation}
\begin{IEEEproof}[\!\!\!\!\!\!\!\!\!\!\! Proof]
  Please refer to Appendix A.
\end{IEEEproof}
It can be seen from (\ref{11221}) that both small-scale and large-scale estimations influence  the communication rate. Large-scale characteristics are reflected in $\mathrm{tr}\left( \boldsymbol{C}_{h,k} \right)$, while small-scale characteristics are evident in both $\rho _{k,n}$ and $\mathrm{tr}\left( \boldsymbol{C}_{h,k} \right)$.
The spectral efficiency (bps/Hz) in $n$-th block for user $k$ can be written as
\begin{equation}
 \begin{split}
   \mathrm{SE}_{k, n}=\log_{2}\left( 1+\gamma _{k,n} \right),\forall k,n.
 \end{split}
 \end{equation}
\section{Problem Formulation}\label{ProblemFormulation}
Considering the dual-scale channel estimation, the system average achievable rate is given by
   \begin{equation}
     \begin{split}
       \label{92035}
       \mathcal{R}= & \ {\frac{1}{N}}\sum_{k=1}^K\sum_{m=1}^M{\!}\sum_{n=1}^{N_m}{\text{SE}_{k,n}}\\
   &-\frac{\left( MT_e+\!\text{mod}\left( T_r,T_b \right) \right)}{NT_b} \sum_{k=1}^K{\text{SE}_{k,1}},
   \end{split}
   \end{equation}
where the second term at the right-hand side of (\ref{92035}) is the rate discount caused by the dual-scale channel estimation, and $\frac{(MT_e + \text{mod}(T_r,T_b))}{NT_b} $  is the time duration occupied by dual-scale channel estimation.  As shown in Fig. \ref{frame},  $M$ is the number of small-scale updates, $N_m$ is the number of blocks per small-scale update, $T_r$ is the time duration of radar sensing, and $T_e$ is the time duration of small-scale estimation.  \par
In this paper, we aim to maximize the average achievable rate given in (\ref{92035}) while guaranteeing the sensing requirements of ISAC users and pure targets by jointly optimizing the dual-scale change estimation-related  time variables, the beamforming vectors of sensing, and communication transmit powers. Thus, the optimization problem can be formulated as
  \begin{subequations}
    \begin{flalign}
      \mathbf{P}1:& \!\mathop{\large \mathrm{max}}\limits_{\scalebox{.6} {\begin{array}{c}
          M, T_r, \left\{ N_m \right\},   \\
          \left\{p_k\right\},\left\{\boldsymbol{W}\!_{r,k}\right\}, \left\{\boldsymbol{W}\!_{r,l} \right\}\\
         \end{array}}} \! \mathcal{R}
        \\
        \!\!\!\!\!\!\!\! \!\!\!\!\!\!\!\!\text{s.t. } & 
          \mathrm{CRB}\left( \theta _k \right) \leqslant \varGamma _{k}, \forall k,\label{91935b}\\
  &\mathrm{CRB}\left( \theta _l \right) \leqslant \varGamma _{l},  \forall l ,\label{91935c}\\ 
  &\sum_{k=1}^K{\mathrm{tr}\left( \boldsymbol{W}_{r,k} \right)}\!+\!\sum_{l=1}^L{\mathrm{tr}\left( \boldsymbol{W}_{r,l} \right)}\leqslant P_{m},\label{91935e}
  \\ 
  &\sum_{k=1}^K{p_k}\leqslant P_{m},\label{91935f}\\
  &\sum_{m=1}^M{N_m=N-}\lfloor \frac{T_r}{T_b} \rfloor,\label{91935d}\\
  & T_r\leqslant NT_b, M\leqslant N,\label{91935g} \\
  &\text{rank}\left( \boldsymbol{W}_{r,k} \right) =1,   \forall k, \text{rank}\left( \boldsymbol{W}_{r,l} \right) =1, \forall l  ,\label{22035h} \\
  & \boldsymbol{W}_{r,k}\succeq 0,  \forall k,  \boldsymbol{W}_{r,l}\succeq 0, \forall l , \label{22035i} 
    \end{flalign}
  \end{subequations}
where $\varGamma _{k}$ and $\varGamma _{l}$ are the maximum detection error requirements for ISAC users and pure targets, respectively. $P_{m}$ is the transmit power budget of BS. Constraints (\ref{91935b}) and (\ref{91935c}) are the basic sensing requirements of the system. Constraints (\ref{91935e}) and (\ref{91935f}) are the total power constraints of large-scale detection and data transmission stages, respectively. Constraint (\ref{91935d}) limits the number of communication transmission blocks to $N-\lfloor \frac{T_r}{T_b} \rfloor $. Constraint    (\ref{91935g}) guarantees that  the times of small-scale and large-scale should be within the scheduling cycle. Constraints (\ref{22035h}) and (\ref{22035i}) represent the basic identities of the radar transmit beamformers.
\section{Problem Decomposition and Transformation}\label{ProblemDecomposition}
The summation of logarithms in the objective function and the complex fractional forms in constraints (\ref{91935b}) and (\ref{91935c}) lead to a non-convex mixed integer nonlinear programming (MINLP) problem. Thus, there is no universal method to solve this problem of highly coupled variables. In the following, we decouple $\mathbf{P}1$ as a time-related subproblem and a power-related resource scheduling subproblem and propose corresponding algorithms to efficiently solve them.
\subsection{Time-Related Optimal Design}
\subsubsection{Determination of the Number of Blocks in Each Update (i.e., $\left\{ N_m \right\}$)}
Given the values of $\left\{p_k\right\},\left\{\boldsymbol{W}\!_{r,k}\right\}$, and $\left\{\boldsymbol{W}\!_{r,l}\right\}$, the second row of (\ref{92035}) is constant and subproblem~$\mathbf{P}2$ for the optimization of $N_m$ can be reformulated as
\begin{subequations}
  \label{92637}
  \begin{flalign}
    \mathbf{P}2: \mathop{\max }\limits_{\left\{ N_m \right\}}\quad & {\frac{1}{N}}\sum_{m=1}^M{\!}\sum_{n=1}^{N_m}{{\varPhi_n}}\\
    \ \ \text{s. t.  }  \quad &\lfloor\frac{T_r}{T_b} \rfloor +\sum_{m=1}^M{N_m=N},\\ 
&1\leqslant N_m, \forall m,
  \end{flalign}
\end{subequations}
where $\varPhi _n=\sum_{k=1}^K{\mathrm{SE}_{k,n}}$ is only effected by index $n$.  Since the discrete scheduling problem is quite intractable, we leverage the following proposition to obtain closed-form solutions.\par
\emph{Proposition 2}: 
The closed-form solutions of $N_m$ in $\mathbf{P}2$ are given by
\begin{small}
\begin{equation}
  \label{92738}
    \begin{split} 
     N_m\in \left\{ \lfloor \frac{N_t}{M} \right. \rfloor ,\;\left. \lfloor \frac{N_t}{M}\rfloor +1 \right\}, \forall m,
    \end{split}
\end{equation}
\end{small}where $N_t=\sum_{m=1}^M{N_m}$. After evenly allocating  $\lfloor \frac{N_t}{M} \rfloor$ in each update,  the  remaining $\text{mod} (N_t, M)$ blocks need to spread evenly in random $\text{mod} (N_t, M)$  updates, which leads to $\lfloor \frac{N_t}{M} \rfloor+1$ blocks, and the other $M-\text{mod} (N_t, M)$ updates  keep to have  $\lfloor \frac{N_t}{M} \rfloor$ blocks.
\begin{IEEEproof}[\!\!\!\!\!\!\!\!\!\!\! Proof]
  Please refer to Appendix B.
\end{IEEEproof}
\vspace{0.5em}
\subsubsection{Determination of the Number of Small-Scale Updates (i.e., $M$) and the Large-Scale Sensing Duration (i.e., $T_r$)}
Based on the results in (\ref{92738}), the achievable rate in (\ref{92035}) can be rewritten as
\begin{small}
  \begin{equation}
    \begin{split} 
      \label{101160}
      \mathcal{R}\!=& {\frac{M}{N}}\sum_{k=1}^K{\sum_{n=1}^{\lfloor \frac{N_t}{M} \rfloor}{\mathrm{SE}_{k,n}}}\!+\! {\frac{1}{N}}\mathrm{mod}\left( N_t,M \right)\! \sum_{k=1}^K{\mathrm{SE}_{k,\lfloor \frac{N_t}{M} \rfloor +1}}\\
      &-\frac{\left( MT_e+\!\text{mod}\left( T_r,T_b \right) \right)}{NT_b}\sum_{k=1}^K{\mathrm{SE}_{k,\!1}}.
    \end{split}
    \end{equation}
\end{small}
Since the objective function is highly complicated, we decouple it by the following steps. For any positive
semidefinite matrices $A$ and $B$, it is  well-known that $\mathrm{tr}\left( AB \right) \leqslant\mathrm{tr}\left( A\right) \mathrm{tr}\left( B \right)$. Thus, based on (\ref{11221}), the lower bound of $\mathrm{SE}_{k,n}$ can be 
\begin{equation}
  \begin{split} 
    \label{101039}
    \gamma _{k,n}^{l}=\frac{p_k\rho _{k,n}^{2}\mathrm{tr}\left( \boldsymbol{C}_{h,k} \right)}{\beta _kP_m \mathrm{tr}\left( \boldsymbol{R}_{k}^{S} \right) \!+\!\sigma _{c}^{2}},\forall k,n,
\end{split}
\end{equation}
where constraint  (\ref{91935f}) on power allocation  is always active with $\sum_{k=1}^K{p_k=P_m}$ since we can scale $p_k, \forall k$ to improve $\gamma _{k,n}^{l}$\cite{bjornson2013optimal}. To solve the subproblem, we have the following proposition. \par
\emph{Proposition 3}: 
For $\left( h-1 \right) T_b<T_r<hT_b, \forall h\in \left\{ 1,2,...,N \right\} $, when $\lfloor\frac{T_r}{T_b} \rfloor$ is fixed and radar's large-scale sensing accuracy is reasonable, i.e., $\hat{\boldsymbol{R}}_k=\beta _k\boldsymbol{R}_{k}^{S}-\boldsymbol{R}_{r,k}\succeq 0$, the derivative of average achievable rate monotonically decreases and then it is segmented concave on $T_r$. \par
\begin{IEEEproof}[\!\!\!\!\!\!\!\!\!\!\! Proof]
  Please refer to Appendix C.
\end{IEEEproof}
According to Proposition 3, the time-related design is a convex problem with fixed $M$ and blocks occupied by $T_r$, i.e., $\lfloor\frac{T_r}{T_b} \rfloor$. Thus, the optimal solution of each segmented problem can be derived via golden search  over fixed $M$ and $\lfloor\frac{T_r}{T_b} \rfloor$. 
\subsection{Power-Related Optimal Design}
\subsubsection{Optimization of the Radar Transmit Beamformers (i.e., $ \left\{\boldsymbol{W}\!_{r,k}\right\}$ and  $\left\{\boldsymbol{W}\!_{r,l} \right\}$}
Given the values of $M, T_r$, and $\left\{ N_m \right\}$,  the optimization of these beamformers is only related to  $\mathrm{SE}_{k,n}$. The problem $\mathbf{P}1$ can be reformulated as 
\begin{subequations}
  \begin{flalign}
    \mathbf{P}3-1:&\!\mathop{\large \mathrm{max}} \limits_{\scalebox{.6} {\begin{array}{c}
      \left\{\boldsymbol{W}\!_{r,k}\right\}, \left\{\boldsymbol{W}\!_{r,l} \right\}\\
       \end{array}}} \! \sum_{k=1}^K{\!}\sum_{n=1}^{\lfloor \frac{N_t}{M} \rfloor +1}{\chi _{n}\,\,\!\mathrm{SE}_{k,n}}\!  \label{3541a}
      \\
      &\quad \ \text{s. t. } \mathrm{constraints}
      \text{ (\ref{91935b})-(\ref{91935e}), (\ref{22035h}) and (\ref{22035i})} \nonumber,
\end{flalign}
\end{subequations}
where $\chi _{n}$ is the average duration of  $n$-th block in $M$ updates, which is a constant in this subproblem, i.e., $\chi _1=\frac{\left( MT_b-MT_e-\text{mod}\left( T_r,T_b \right) \right)}{NT_b}$ obtained from (\ref{101160}). The deep coupling of variables in both the objective function and constraints makes the problem challenging. We resolve it through the following steps.\par
Firstly, through Lagrangian Dual Transform\cite{shen2018fractional},  the sum of logarithmic form  of $\mathrm{SE}_{k,n}$ can be decoupled by introducing a new set of variables $\boldsymbol{\varsigma }=\left[ \varsigma _{1,1},\dots ,\varsigma _{K,\lfloor \frac{N_t}{M} \rfloor +1} \right]$ as 
\begin{equation}
\begin{split} 
  \label{101742}
\mathrm{SE}_{k,n}&\left( \boldsymbol{W}\!_{r,k},\varsigma _{k,n} \right)\\ &\!\!\!\!\!\!\!\!\!\!\!\!\!\!\!\!=\chi _{n}\!\left(\mathrm{ln}2\!-\!\frac{\mathrm{ln}2(1+\varsigma _{k,n})}{\kappa _{k,n}\mathrm{tr}\left( \boldsymbol{C}_{h,k} \right)+\!1\!}\!+\!\log_2 \left( 1+\varsigma _{k,n} \right)\! \right), \forall k, n,
\end{split}
\end{equation}
where $\kappa _k=\frac{p_k\rho _{k,n}^{2}}{\beta _kP_m\mathrm{tr}\left( \boldsymbol{R}_{k}^{S} \right)\!+\!\sigma _{c}^{2}}$.
With the above equivalent reformulation, we can interactively optimize the variables in $\mathrm{SE}_{k,n}\left( \boldsymbol{W}\!_{r,k},\varsigma _{k,n} \right)$. When $\boldsymbol{W}\!_{r,k}$ is fixed, the optimal $\varsigma _{k,n}$ can be obtained by setting $\partial  \mathrm{SE}_{k,n}\left( \boldsymbol{W}\!_{r,k},\varsigma _{k,n} \right) /\partial \varsigma _{k,n}=0$, i.e., $\varsigma _{k,n}=\kappa _{k,n}\mathrm{tr}\left( \boldsymbol{C}_{h,k} \right)$. When $\varsigma _{k,n}$ is fixed, the optimization of $\boldsymbol{W}\!_{r,k}$ has the sticky sum-of-ratios objective function. Thus, we propose the following theorem to obtain an equivalent objective function in subtractive form, which has the same allocation policy as the original problem.\par
\emph{Theorem 1}: 
If {\small $\left\{ \boldsymbol{W}\!_{r,k}^{*} \right\}$} is the optimal solution of  $\mathbf{P}3-1$,  there exist two vectors {\small$\boldsymbol{\varpi }^*=\left[ \varpi _{1,1}^{*},\dots ,\varpi _{K,\lfloor \frac{N_t}{M} \rfloor +1}^{*} \right]$}, and {\small$\boldsymbol{\vartheta }^*=\left[ \vartheta _{1,1}^*,\dots ,\vartheta _{K,\lfloor \frac{N_t}{M} \rfloor +1}^{*} \right]$} that make  {\small $\left\{ \boldsymbol{W}\!_{r,k}^{*} \right\}$} satisfying the Karush-Kuhn-Tucker (KKT) conditions of the following problem
\begin{small}
  \begin{subequations}
    \begin{flalign}
     \!\!\! \mathop{\mathrm{min}} \limits_{\scalebox{.55} {\begin{array}{c}
        \boldsymbol{W}\!_{r,k}\in \mathcal{F} 
         \end{array}}}& \!\!\!  \sum_{k=1}^K\!\!\!\sum_{n=1}^{\lfloor \frac{N_t}{M} \rfloor +1}{{\!\!\!\!\!\varpi _{k,n}^{*}\!\!\left[\chi _{n}\mathrm{ln}2\left( 1 \!+\varsigma _{k,n} \right)\! -\!\vartheta _{k,n}^{*}\!\left(\kappa _{k,n}\mathrm{tr}\left( \boldsymbol{C}_{h,k} \right) \!+\!1 \right) \right]}}, \tag{38}
  \end{flalign}
  \end{subequations}
\end{small}where $\mathcal{F}$ is the feasible solution set of $\mathbf{P}3-1$, and the constant part $\sum_{k=1}^K{\!}\sum_{n=1}^{\lfloor \frac{N_t}{M} \rfloor +1}{\,\,\!}\chi _{n}\left( 1+\!\log \left( 1+\varsigma _{k,n} \right) \right)$ from (\ref{101742}) is dropped here. Besides, {\small $\boldsymbol{W}\!_{r,k}^{*}$} needs to satisfy the following equations
\begin{equation}
  \begin{split} 
    \label{12244}
   &\vartheta _{k,n}^{*}\left( \kappa _{k,n}\mathrm{tr}\left( \boldsymbol{C}_{h,k} \right) +\!1 \right) -\chi _{n}\mathrm{ln}2\left( 1+\varsigma _{k,n} \right) =0,  \forall k,n. \\
   &\varpi _{k,n}^{*}\left( \kappa _{k,n}\mathrm{tr}\left( \boldsymbol{C}_{h,k} \right) +\!1 \right)-1=0,  \forall k,n.
  \end{split}
  \end{equation}
\par
\begin{IEEEproof}[\!\!\!\!\!\!\!\!\!\!\! Proof]
  Please refer to \cite{bai2020latency}.
\end{IEEEproof}
Therefore, the sum-of-ratios form in (\ref{101742}) can be transformed to an equivalent subtractable form, and the newly introduced variables can be alternately optimized with the original variables.\par The optimal solutions of  $\boldsymbol{\varpi }$ and $\boldsymbol{\vartheta}$ can be obtained with the well-known damped Newton's method when the convergence is achieved. For notational simplicity, we define 
\begin{subequations}
  \begin{flalign}
    &\check{\varphi}_{k,n}\!\left( \vartheta _{k,n} \right)\! =\vartheta _{k,n}\!\left( \kappa _{k,n}\mathrm{tr}\!\left( \boldsymbol{C}_{h,k} \right) \!+\!1 \right)\!-\!\chi _{n}\ln 2\left( 1\!+\varsigma _{k,n} \right),\nonumber \\& \qquad \qquad\qquad\qquad\qquad\qquad\qquad\qquad\qquad\quad \!\!\!\!\!\!\!\!\!\! \forall k,n,\\
   & \hat{\varphi}_{k,n}\!\left( \varpi _{k,n} \right) \!=\varpi _{k,n}\!\left( \kappa _{k,n}\mathrm{tr}\left( \boldsymbol{C}_{h,k} \right)\! +\!1 \right) \!-\!1, \ \forall k,n.
    \end{flalign}
  \end{subequations}Then, in $t$-th iteration,  $\varpi _{k,n}^{t+1}$ and $\vartheta _{k,n}^{t+1}$ can be updated as\begin{subequations}
    \begin{flalign}
      &\varpi _{k,n}^{t+1}=\varpi _{k,n}^{t}-\frac{\mathfrak{w} ^{x^{\left( t+1 \right)}}\hat{\varphi}_{k,n}\left( \varpi _{k,n}^{t} \right)}{\kappa _{k,n}^{t+1}\mathrm{tr}\left( \boldsymbol{C}_{h,k}^{t+1} \right) +\!1},\quad\forall k,n,
      \\
     &\vartheta _{k,n}^{t+1}=\vartheta_{k,n}^{t}-\frac{\mathfrak{w} ^{x^{\left( t+1 \right)}}\check{\varphi}_{k,n}\left( \vartheta _{k,n}^{t} \right)}{\kappa _{k,n}^{t+1}\mathrm{tr}\left( \boldsymbol{C}_{h,k}^{t+1} \right) +\!1},\!\!\!\quad\quad\forall k,n,
      \end{flalign}
    \end{subequations} where $\mathfrak{w} \in \left( 0,1 \right)$ and $x^{\left( t+1 \right)}\in \left\{ 1, 2,3,... \right\}$  is the smallest integer and they satisfy the following condition\cite{bai2020latency}
    \begin{small}
      \begin{equation}
        \begin{split} 
          \sum_{k=1}^K&{\!}\!\!\sum_{n=1}^{\lfloor \frac{N_t}{M} \rfloor +1}{\left| \hat{\varphi}_{k,n}\left( \varpi _{k,n}^{t}\!-\!\frac{\mathfrak{w} ^x\hat{\varphi}_{k,n}\left( \varpi _{k,n}^{t} \right)}{\kappa _{k,n}^{t+1}\mathrm{tr}\left( \boldsymbol{C}_{h,k}^{t+1} \right) +\!1} \right) \right|^2}\\
         &+\sum_{k=1}^K{\!}\!\!\sum_{n=1}^{\lfloor \frac{N_t}{M} \rfloor +1}{\left| \check{\varphi}_{k,n}\left( \vartheta _{k,n}^{t}\!-\!\frac{\mathfrak{w} ^x\check{\varphi}_{k,n}\left( \vartheta _{k,n}^{t} \right)}{\kappa _{k,n}^{t+1}\mathrm{tr}\left( \boldsymbol{C}_{h,k}^{t+1} \right) +\!1} \right) \right|^2}\\
         &\leqslant \left( 1-\epsilon \mathfrak{w} ^x \right) ^2\sum_{k=1}^K{\!}\!\!\sum_{n=1}^{\lfloor \frac{N_t}{M} \rfloor +1}\!\!{\left[ \left| \hat{\varphi}_{k,n}\left( \varpi _{k,n}^{t} \right) \right|^2\!+\!\left| \check{\varphi}_{k,n}\left( \vartheta _{k,n}^{t} \right) \right|^2 \right]},
        \end{split}
        \end{equation}
    \end{small}where $\epsilon \in \left( 0,1 \right)$, and the damped
  Newton method converges to the optimal solutions until the equations in (\ref{12244}) are achieved.\par
  We now focus on the optimization of $\left\{\boldsymbol{W}\!_{r,k}\right\}$ and $\left\{\boldsymbol{W}\!_{r,l} \right\}$. To this end, each term of the objective function for optimization of radar transmit beamformers can be simplified as $-\kappa _{k,n}\varpi _{k,n}^{*}\vartheta _{k,n}^{*}\mathrm{tr}\left( \boldsymbol{C}_{h,k} \right)$. Combining the linear relaxation in Appendix C, we have
  \begin{equation}
    \begin{split} 
        \label{112572}
        \mathrm{tr}\left( \boldsymbol{C}_{h,k} \right) \!\geqslant & \ \mathcal{U} \left( \mathrm{CRB}^{\left( t \right)}\left( \theta _k \right) \right) \mathrm{CRB}\left( \theta _k \right) +\mathrm{tr}\!\left( \boldsymbol{C}_{h,k}^{\left( t \right)} \right)\!\!\\& -\mathcal{U} \!\left(\! \mathrm{CRB}^{\left( t \right)}\!\left( \theta _k \right) \!\right) \mathrm{CRB}^{\left( t \right)}\!\left( \theta _k \right)\triangleq \mathcal{U} _{lb,k}^{\left( t \right)},\forall k,
    \end{split}
    \end{equation}
where  $\mathcal{U} \left( \mathrm{CRB}^{\left( t \right)}\left( \theta _k \right) \right) <0$. Through dropping the constant parts, the subproblem on radar transmit beamformers can be reformulated as 
\begin{subequations}
  \begin{flalign}
    \mathbf{P}3-2:& \!\mathop{\large \mathrm{min}}\limits_{\scalebox{.6} {\begin{array}{c}
      \left\{\boldsymbol{W}\!_{r,k}\right\}, \left\{\boldsymbol{W}\!_{r,l} \right\}\\
       \end{array}}} \sum_{k=1}^K{\!}\tilde{\beta}_k\mathrm{CRB}\left( \theta _k \right) \!
      \\
      &\quad \ \text{s. t. } \mathrm{constraints}  \text{ (\ref{91935b})-(\ref{91935e}), (\ref{22035h}) and (\ref{22035i})},\nonumber
\end{flalign}
\end{subequations}
where $\tilde{\beta}_k=-\mathcal{U} \left( \mathrm{CRB}^{\left( t \right)}\left( \theta _k \right) \right)\sum_{n=1}^{\lfloor \frac{N_t}{M} \rfloor +1}{\kappa _{k,n}\varpi _{k,n}^{*}\vartheta _{k,n}^{*}}$ is the positive weighted constant.  By introducing a set of auxiliary variables $\boldsymbol{\mathfrak{t}}=\left[ \mathfrak{t} _1,\mathfrak{t} _2,\dots, \mathfrak{t} _K \right]$, $\mathbf{P}3-2$ can be equivalently reformulated as 
\begin{subequations}
  \begin{flalign}
    \mathbf{P}3-3:& \mathop{\large \mathrm{min}} \limits_{\scalebox{.788} {\begin{array}{c}
      \left\{\boldsymbol{W}\!_{r,k}\right\}, \left\{\boldsymbol{W}\!_{r,l} \right\}\\
      \left\{\mathfrak{t} _k\right\} \end{array}}}\sum_{k=1}^K\tilde{\beta}_k\mathfrak{t} _k\!
      \\
      \text{s. t. } \mathrm{tr}& {\footnotesize\left( \dot{\boldsymbol{A}_k}\boldsymbol{W}_{r,k}\dot{\boldsymbol{A}}_k^H \right)} -{\footnotesize\frac{\left| \mathrm{tr}\left( \boldsymbol{A}_k\boldsymbol{W}_{r,k}\dot{\boldsymbol{A}}_k^H \right) \right|^2}{\mathrm{tr}\left( \boldsymbol{A}_k\boldsymbol{W}_{r,k}\boldsymbol{A}_k^H \right)}} \nonumber \\ &\geqslant {\small \frac{\sigma _{r}^{2}}{2T_r\left| \dot{\alpha}_k \right|^2}\,\,\max \left\{ \frac{1}{\mathfrak{t} _k},\frac{1}{\varGamma _{k}} \right\}}, \forall k, \label{101846b}\\
     & {\footnotesize\mathrm{tr}\left( \dot{\boldsymbol{A}_l}\boldsymbol{W}_{r,l}\dot{\boldsymbol{A}}_l^H \right) -\frac{\left| \mathrm{tr}\left(\boldsymbol{A}_{l}\boldsymbol{W}_{r,l}\dot{\boldsymbol{A}}_l^H \right) \right|^2}{\mathrm{tr}\left(\boldsymbol{A}_{l}\boldsymbol{W}_{r,l}\boldsymbol{A}_l^H \right)}}\nonumber \\ &\geqslant {\small\frac{\sigma _{r}^{2}}{2T_r\varGamma _{l}\left|\dot{\alpha}_l \right|^2}},  \forall k,\label{101846c}
     \\
     & \mathfrak{t}_k\geqslant 0,   \forall k, \label{101846d}
     \\
    &\mathrm{constraints} \  \text{(\ref{91935e}), (\ref{22035h}) and (\ref{22035i}), }\nonumber
\end{flalign}
\end{subequations}
where constraint (\ref{101846b}) is the combination of constraint (\ref{91935b}) and  $\tilde{\beta}_k\mathrm{CRB}\left( \theta _k \right) \leqslant \mathfrak{t} _k$, which also indicates $\mathfrak{t}_k\geqslant 0$ in constraint (\ref{101846d}). Constraint (\ref{101846c}) is the equivalent transformation of (\ref{91935c}). According to Schur's complement\cite{song2023intelligent}, constraints (\ref{101846b}) and (\ref{101846c}) can be equivalently rewritten as the following semi-definite forms
\begin{subequations}
  \begin{flalign}
&\left[ \begin{matrix}
	\mathrm{tr}\left( \dot{\boldsymbol{A}_k}\boldsymbol{W}_{r,k}\dot{\boldsymbol{A}}_k^H \right)-\bar{\mathfrak{t}}_k&		\!\!\!\mathrm{tr}\left( \boldsymbol{A}_k\boldsymbol{W}_{r,k}\dot{\boldsymbol{A}}_k^H \right)\\
	\mathrm{tr}\left( \dot{\boldsymbol{A}_k}\boldsymbol{W}_{r,k}\boldsymbol{A}_k^H \right)&		\!\!\! \mathrm{tr}\left( \boldsymbol{A}_k\boldsymbol{W}_{r,k}\boldsymbol{A}_k^H \right)\\
\end{matrix} \right] \succeq 0,\forall k,\label{101847a}
\\
&\left[ \begin{matrix}
	\mathrm{tr}\!\left( \!\dot{\boldsymbol{A}_l}\boldsymbol{W}_{r,l}\dot{\boldsymbol{A}}_l^H \right)\! -\!\frac{\sigma _{r}^{2}}{2T_r\varGamma _{l}\left| \dot{\alpha}_l \right|^2}&		\!\!\! \mathrm{tr}\!\left( \! \boldsymbol{A}_{l}\boldsymbol{W}_{r,l}\dot{\boldsymbol{A}}_l^H \right)\\
	\mathrm{tr}\left( \dot{\boldsymbol{A}_l}\boldsymbol{W}_{r,l}\boldsymbol{A}_l^H \right)&		\!\!\! \mathrm{tr}\left(\boldsymbol{A}_{l}\boldsymbol{W}_{r,l}\boldsymbol{A}_l^H \right)\\
\end{matrix} \right] \!\! \succeq 0, \forall l,\label{101847b}
\end{flalign}
\end{subequations}
where $\bar{\mathfrak{t}}_k=\frac{\sigma _{r}^{2}}{2T_r\left| \dot{\alpha}_k \right|^2}\,\,\max \left\{ \frac{1}{\mathfrak{t} _k},\frac{1}{\varGamma _{k}} \right\}$. On the left side of the greater-than sign, $-\bar{\mathfrak{t}}_k$ is an obviously concave function with respect to $\mathfrak{t}_k$.   Then, relaxing the non-convex rank-one constraint in (\ref{22035h}), $\mathbf{P}3-3$ can be reformulated as the following semidefinite relaxation (SDR) problem
\begin{subequations}
  \begin{flalign}
    \mathbf{P}3-4:& \mathop{\large \mathrm{min}} \limits_{\scalebox{.788} {\begin{array}{c}
      \left\{\boldsymbol{W}\!_{r,k}\right\}, \left\{\boldsymbol{W}\!_{r,l} \right\}\\
      \left\{\mathfrak{t} _k\right\}\end{array}}}\sum_{k=1}^K\mathfrak{t} _k\!
      \\
      \text{s. t. } 
    &\mathrm{constraints}\  \text{(\ref{91935e}), (\ref{22035i}),(\ref{101847a}), (\ref{101847b}), and (\ref{101846d}). \nonumber}
\end{flalign}
\end{subequations}
It can be observed that $\mathbf{P}3-4$ is a convex problem, which can be efficiently solved by existing optimization solvers. However, solutions of the relaxed problem typically do not satisfy the rank-one condition, i.e., $\text{rank}\left( \boldsymbol{W}_{r,k} \right)\ne 1$. The Gaussian randomization method is adopted to construct the rank-one solutions from the optimal higher-rank solutions\cite{8811733}.
\subsubsection{Optimization of the Communication Transmit Powers, (i.e., $\left\{ p_k \right\}$)}
Since the subtraction part of (\ref{92035}) can be transformed as the form in (\ref{3541a}), i.e., $\chi _1$, $p_k$  is obviously concave for the power optimization problem, and optimal solutions of this convex problem can be efficiently obtained.

\subsection{Proposed Algorithm}
In summary, we propose a segmented golden search-based joint optimization method to address the original problem $\mathbf{P}1$ by alternately optimizing two parts: time-related optimal design and power-related optimal design. The pseudocode of the proposed method is given in Algorithm 1.\par

The computational complexity of the golden search-based joint optimization algorithm is analyzed as follows. The complexity of time-related optimal design is dominated by the golden search part in Steps 5-20 with the total computational complexity  {\small$\mathcal{O} \left( \frac{\left( N-N_{s}^{\min} \right) \left( N-N_{s}^{\min}+1 \right)}{2}\log _2\left( \frac{T_b}{\varepsilon} \right) \right)$}. The complexity of power-related optimal design is mainly dependent on the interior point method used by CVX,  since the other steps have explicit mathematical expressions. The computational complexities for updating radar transmit beamformers $\left\{\boldsymbol{W}\!_{r,k}\right\}$, $\left\{\boldsymbol{W}\!_{r,l}\right\}$ and communication transmit
power $\left\{p_k\right\}$ are {\small$\mathcal{O} \left( \left( \left( K+L \right) L_{t}^{2}+K \right) ^{3.5} \right)$} and {\small$\mathcal{O} \left( K^{3.5} \right)$}, respectively. Thus, in each iteration, the computational complexity of Algorithm 1 is {\small$\mathcal{O}\! \! \left( \!\max\! \left\{\! \frac{\left( N\!-\!N_{s}^{\min} \right) \left( N-N_{s}^{\min}\!+1 \right)}{2}\log _2\!\left( \frac{T_b}{\varepsilon} \right)\! ,\! \left(\! \left( K\!\!+\!L \right)\! L_{t}^{2}\!\!+\!\!K \right) ^{3.5}\! \right\} \!\right)$.}
\vspace{0.5em}
\begin{algorithm}
  \caption{Segmented Golden Section Search-based Joint Optimization Method for Problem $\mathbf{P}1$ }\label{alg:1}
  \begin{algorithmic}[1]
  \STATE \textbf{Initialize}: $\left\{p_k\right\},\left\{\boldsymbol{W}\!_{r,k}\right\}, \left\{\boldsymbol{W}\!_{r,l} \right\}$, threshold $\varepsilon$, $C_{th}$, and golden ratio $\phi \approx 0.618$.
  \STATE \textbf{Repeat}
  \STATE \textbf{\underline{Time-Related Optimal Design}}
  \STATE Calculate the minimum blocks $N_s^{\mathrm{min}}$ required to satisfy sensing requirements in (\ref{91935b}) and (\ref{91935c}).
  \FOR{$\lfloor\frac{T_r}{T_b} \rfloor\!=\! N_s^{\mathrm{min}}\!:\!1\!:\!N-1$ }
    \FOR{$M \!= \!1\!:\!1\!:\!N-\lfloor\frac{T_r}{T_b}\rfloor$}
      \STATE Calculate $\left\{ N_m \right\}$ in each segmented $\lfloor\frac{T_r}{T_b} \rfloor$ according to \emph{Proposition 2}.
      \STATE Set the upper bound as $\hat{T}^{up}=\lfloor\frac{T_r}{T_b} \rfloor+T_b$, the lower bound as $\hat{T}^{lb}=\lfloor\frac{T_r}{T_b} \rfloor$.
      \STATE Calculate $T_1 = \hat{T}^{up} - \phi(\hat{T}^{up} - \hat{T}^{lb})$ and $T_2 = \hat{T}^{lb} + \phi(\hat{T}^{up} - \hat{T}^{lb})$.
      \WHILE {$\left| \hat{T}^{up}-\hat{T}^{lb} \right|>\varepsilon$}
        \IF{$\mathcal{R}\left( T_1 \right) < \mathcal{R}\left( T_2 \right)$}
          \STATE Set $\hat{T}^{lb} = T_1$, $T_1 = T_2$, and update $T_2 = \hat{T}^{lb} + \phi(\hat{T}^{up} - \hat{T}^{lb})$.
        \ELSE
          \STATE Set $\hat{T}^{up} = T_2$, $T_2 = T_1$.
          \STATE Update $T_1 = \hat{T}^{up} - \phi(\hat{T}^{up} - \hat{T}^{lb})$.
        \ENDIF 
      \ENDWHILE
      \STATE Obtain the achievable rate in this loop.
    \ENDFOR
  \ENDFOR
  \STATE Obtain $M, T_r, \left\{ N_m \right\}$ that maximizes the achievable rate.
  \STATE \textbf{\underline{Power-Related Optimal Design}}
  \STATE Update $\boldsymbol{\varpi }$ and $\boldsymbol{\vartheta}$ by the damped Newton's method. 
  \STATE Update $\left\{\boldsymbol{W}\!_{r,k}\right\}, \left\{\boldsymbol{W}\!_{r,l} \right\}$ by solving $\mathbf{P}3-4$ and apply Gaussian randomization.
  \STATE Update $\left\{p_k\right\}$ by solving corresponding convex problem.
  \STATE \textbf{Until} the increase of the system communication achievable rate is below a threshold $C_{th}$.
  \end{algorithmic}
\end{algorithm}

\section{Performance Evaluation}\label{PerformanceEvaluation} 
In this section, we first introduce the simulation setup and then provide numerical results to demonstrate the   effectiveness of our proposed scheme.

\subsection{Simulation Setup}
We consider the simulation scenario consisting of a central ISAC BS with five users randomly distributed, and one fixed pure target which is located at direction $45^{\circ}$. The system  operates at a carrier frequency of $f_c=28$ GHz with a power budget of $P_{m}=46$ dBm.   The large-scale signal attenuation at a reference distance of 1 m is set as 30 dB,  and the path-loss exponent is set as 2.2.  We consider that the BS has $L_t=L_r$ transceivers and all users have the same speed\footnotemark{}. However, the speed varies within the range of [0, 60] km/h, leading to a corresponding variation of temporal correlation coefficient at one block interval $\rho _{k,1}$ between [1, 0.9], as derived from (\ref{22416}). Referring to the settings in \cite{chen2023impact}, the variation of 
the  block fading is consistent with  $\rho _{k,n}=\rho _n=\left( \rho _1 \right) ^n$. The symbol duration and the number of symbols in each block are $T_s=2$ us and $Q_b=50$, respectively. The scheduling cycle is 6 ms with $N=60$.  The sensing requirements of all targets are the same, with  $\varGamma _{k}=\varGamma _{l}=\varGamma$.  The $\left(p,q \right)$-th element of $\boldsymbol{R}_{k}^{S}$ in the one-ring model is given
by \cite{li2023impact} 
\begin{equation}
  \begin{split} 
    \left[ \boldsymbol{R}_{k}^{S} \right] _{p,q}=\frac{\beta _k}{2\Delta \theta}\int_{\theta _k-\Delta \theta}^{\theta _k+\Delta \theta}{\exp \left[ j\pi \left( q-p \right) \sin \left( \theta \right) \right]}d\theta, 
\end{split}
\end{equation}
where the  single-side angular spread $\Delta \theta$ is set as $2^{\circ}$. Unless otherwise stated, the other parameters are set as follows: $L_t=L_r=8$, $\rho _1=0.97$ (30km/h), $\varGamma=1$, and $Q_e=15$ with corresponding training energy ratio $\gamma _e=1.5\times 10^6$.\par
\footnotetext{ Although the parameter settings are simplified as  $L_t=L_r$ and all users have the same speed, the proposed analysis and method are valid in any general setting without any modifications.}
To evaluate the performance of our proposed scheme, we compare it with different benchmark schemes as follows.

\begin{itemize}
  \item Single Small-scale Update (SSU): it only conducts one time small-scale update at the beginning of each scheduling cycle's start\cite{chen2023impact}, and the optimization of other variables refers to Algorithm 1.
  \item Frequent Small-scale Update (FSU): it performs a small-scale update at each communication block \cite{gao2022integrated}, and the optimization of other variables refers to Algorithm 1.
  \item Random Block Allocation (RBA): The design of $M$ and $T_r$ is consistent with the proposed algorithm, but the number of blocks allocated for small-scale updates, $\left\{ N_m \right\}$, is randomized.
  \item Equal Power Allocation (EPA): The communication power of BS is evenly distributed to each user and the optimization of other variables refers to Algorithm~1. 
\end{itemize}

\subsection{Convergence Behavior of Proposed Algorithm}
Fig. \ref{cover} shows the convergence performance of our proposed algorithm for different temporal correlation coefficients. The speeds corresponding to $\rho _1=1$, $0.97$, and $0.9$ are  0 km/h, 30 km/h, and 60 km/h, respectively. It can be observed that the system achievable rate  monotonically increases at the first few iterations and converges within 5 iterations. The system utility improves significantly during the first two iterations and subsequent iterations gradually improve until convergence. Since the system achievable rate has an upper bound and is increasing at each iteration, the results in Fig. \ref{cover} validate the convergence of our proposed algorithm.
\begin{figure*}[t]
  \setcounter{figure}{3}
  \centering
  \begin{minipage}[t]{0.47\textwidth}
    \centering
    \includegraphics[width=7.7cm]{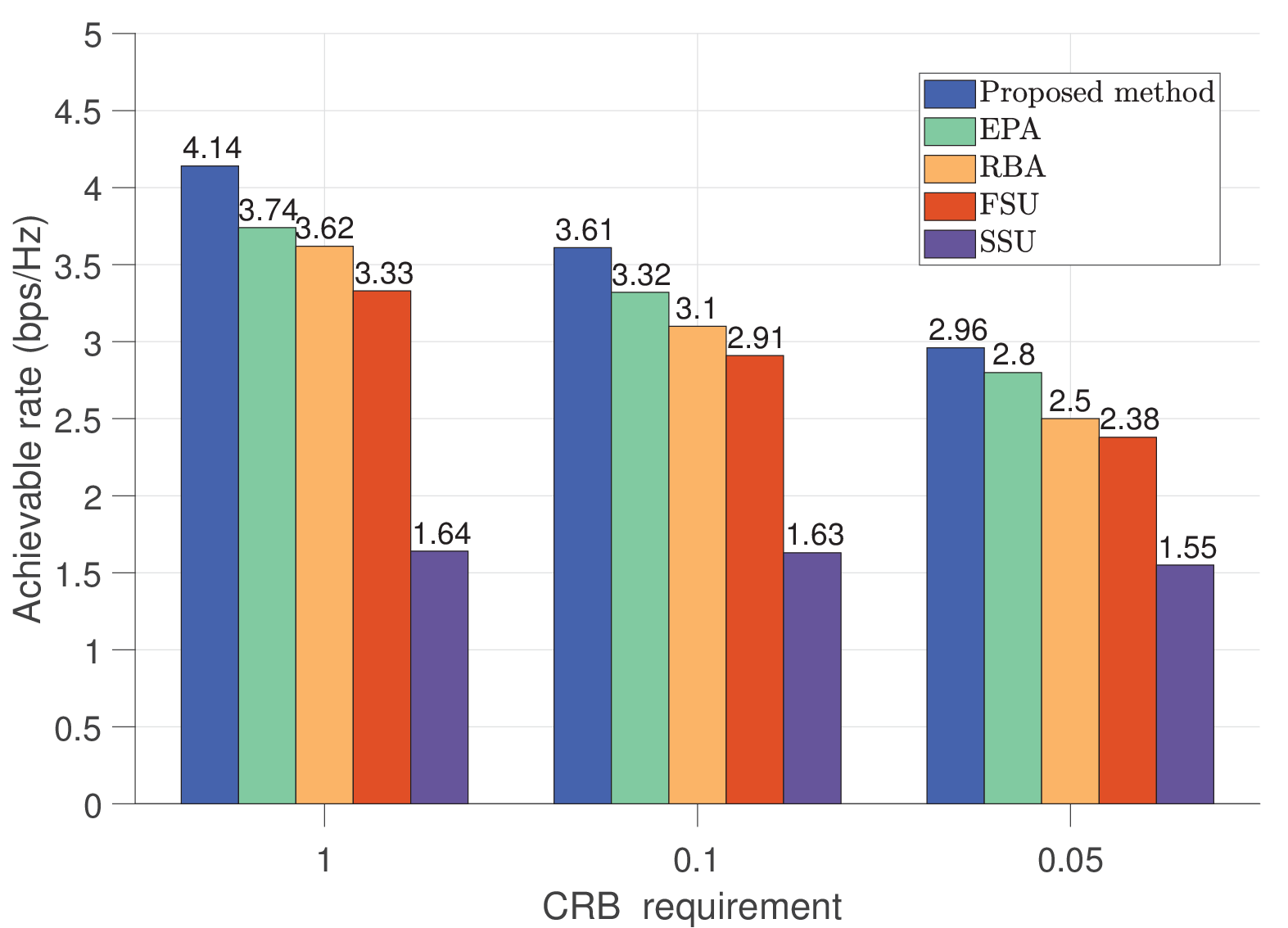}
    \caption{Achievable rate versus the CRB requirement, $\varGamma$.}
    \label{comcrb}
    \end{minipage}
    \begin{minipage}[t]{0.47\textwidth}
      \centering
      \includegraphics[width=7cm]{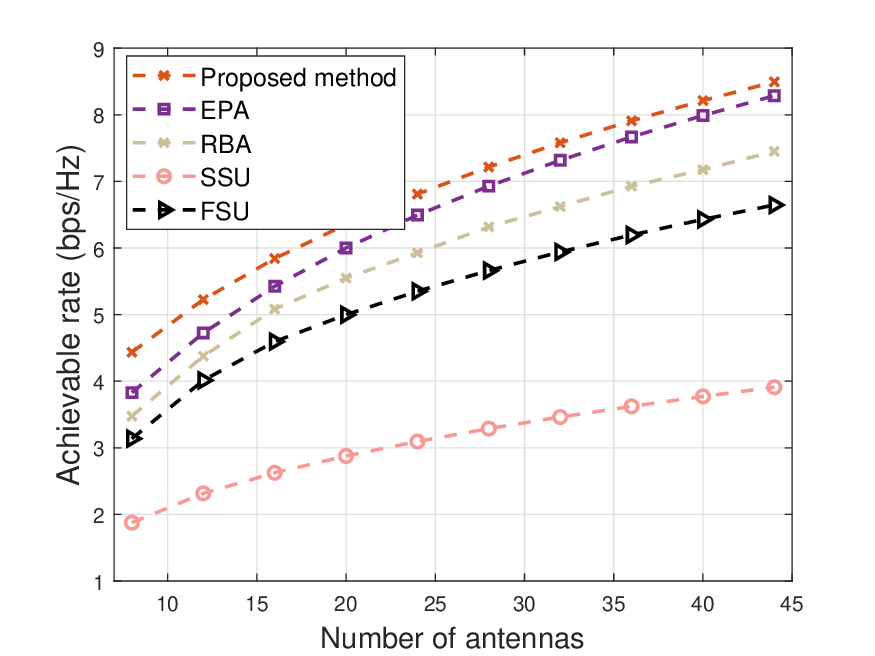}
      \caption{Achievable rate versus the number of antennas, $L_t=L_r$.}
      \label{comanten}
      \end{minipage}
      \vspace{-1em}
  \end{figure*}
\begin{figure}[h]
  \setcounter{figure}{2}
  \centerline{\includegraphics[width=7cm]{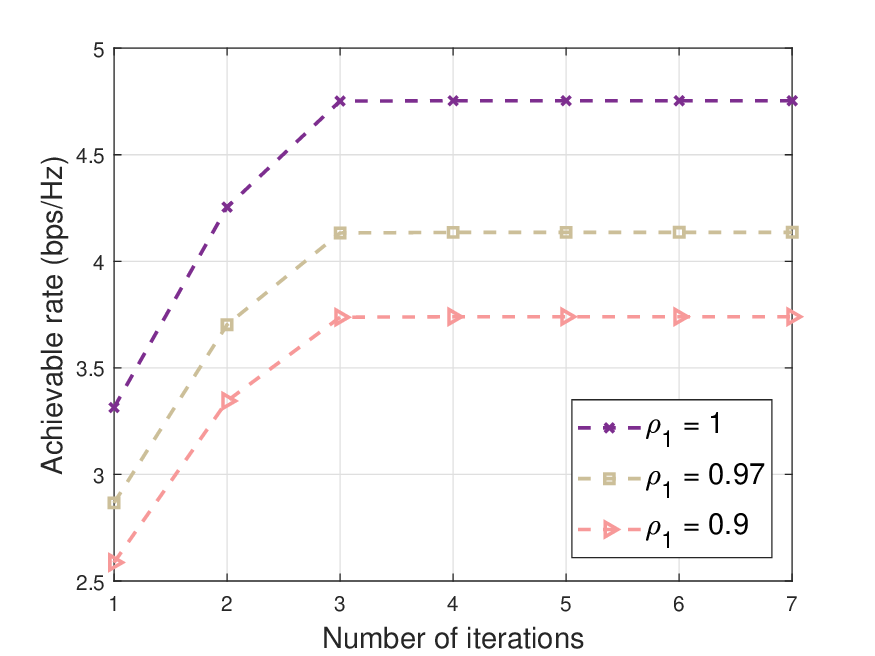}}
  \caption{Convergence behavior of our proposed algorithm.}
  \label{cover}
  \vspace{-1.3em}
\end{figure}

\begin{figure*}[h]
  \setcounter{figure}{5}
  \centering
      \begin{minipage}[t]{0.47\textwidth}
        \hspace{1.4em}
      \includegraphics[width=7cm]{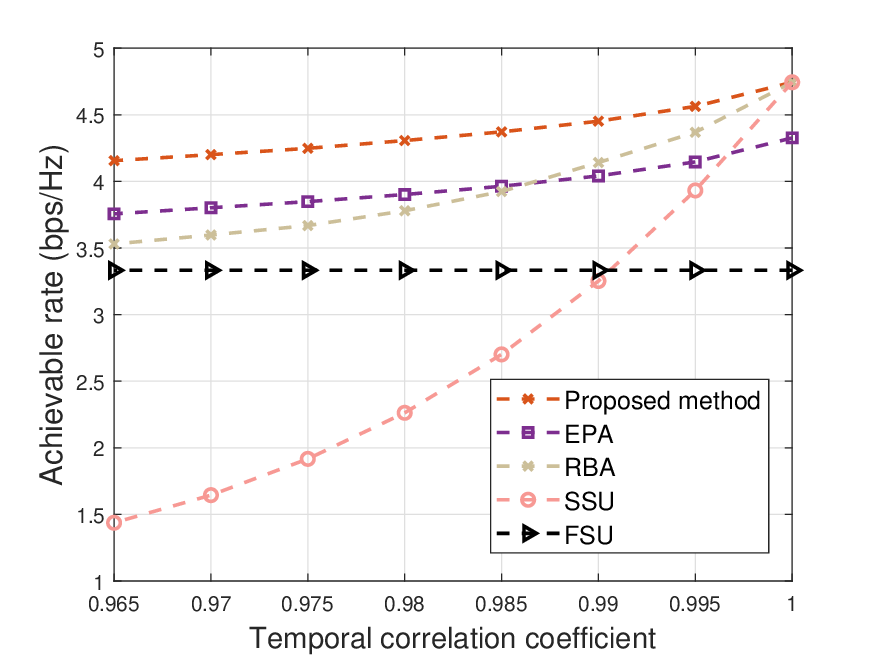}
      \caption{Achievable rate versus the temporal correlation coefficient, $\rho _1$.}
      \label{comrou}
      \end{minipage} \begin{minipage}[t]{0.47\textwidth}
        \hspace{1.4em}
        \includegraphics[width=7cm]{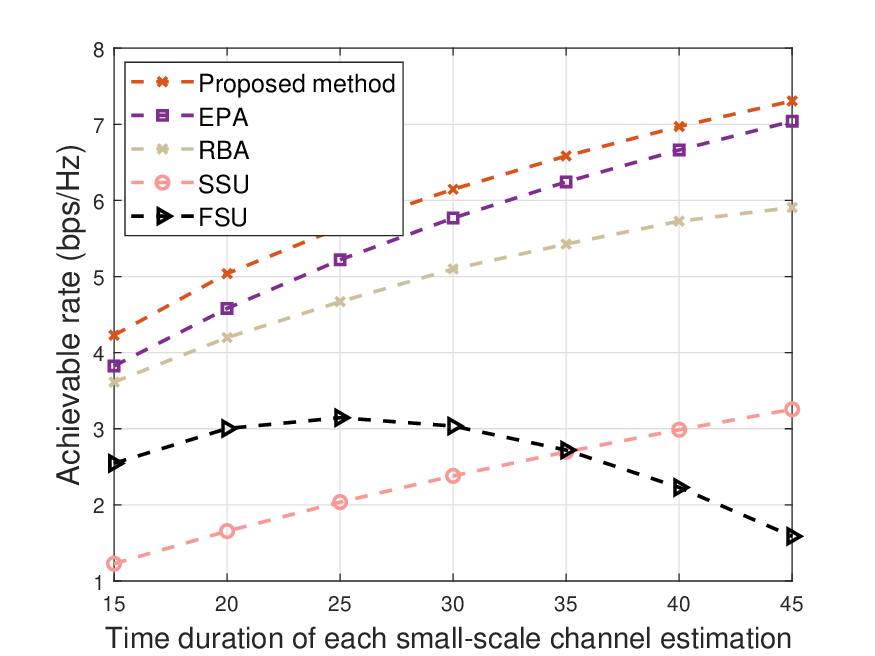}
        \caption{Achievable rate versus the time duration of each small-scale channel estimation, $Q_e$.}
        \label{compayload}
      \end{minipage}
      \vspace{-1em}
  \end{figure*}
\subsection{Performance Comparison}
\subsubsection{Performance with different CRB Requirement, $\varGamma$, and  the number of antennas, $L_t=L_r$}
Fig. \ref{comcrb} shows the achievable rate versus the CRB requirement $\varGamma$ with different schemes, where a lower value of $\varGamma$ means more accurate sensing requirement. It can be observed that our proposed method outperforms the other four counterparts in terms of the highest achievable rate. Specifically, when $\varGamma=1$, our proposed method improves the achievable rate by 10.7\%, 14.4\%, 24.3\%, and 152.4\%  compared to the schemes of ``EPA'', ``RBA'', ``FSU'', and ``SSU'', respectively. Meanwhile,  Fig. \ref{comcrb} demonstrates that the achievable rates of all schemes decrease with more accurate sensing requirements (i.e., a lower value of $\varGamma$). This is because high sensing accuracy requirements conflict with the goal of improving the achievable rate, as sensing consumes a considerable amount of time resources. Fig.~\ref{comanten} depicts the achievable rate with different numbers of antennas. Since a larger number of antennas provides greater beamforming gains, it is observed that the achievable rate increases with the increasing number of antennas.  Moreover, as the number of antennas increases,  the improved antenna gain enhances communication for users with weaker channel conditions, thus driving the performance of the
proposed communication power allocation closer to EPA. Concretely, when the number of antennas is 8 and 44,  the proposed method achieves 15.9\% and 2.5\% performance improvements over ``EPA'', respectively.
\begin{figure*}[t]
  \centering
  \begin{minipage}[t]{0.47\textwidth}
    \centering
    \includegraphics[width=7.3cm]{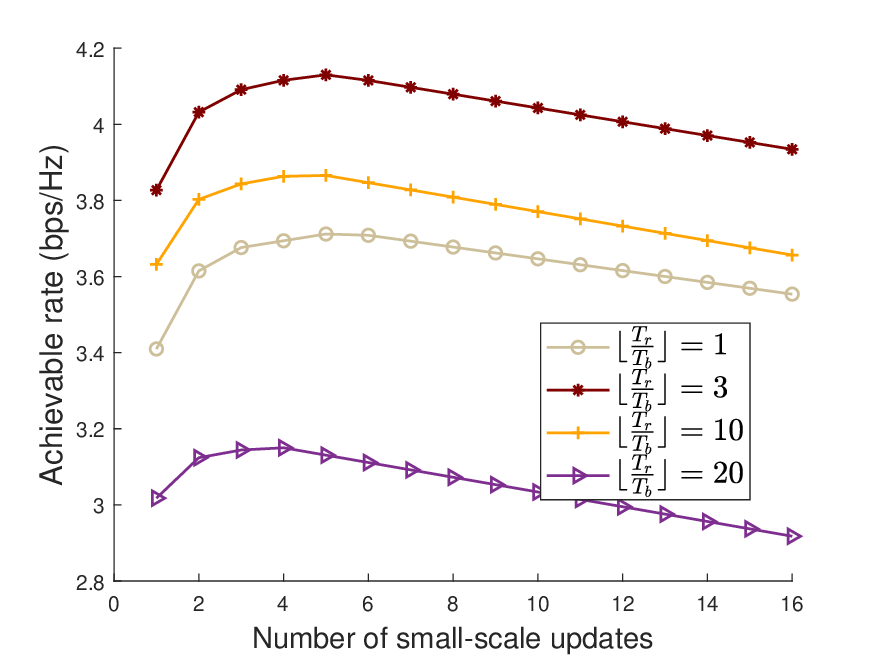}
    \caption{Achievable rate  versus the number of\\   small-scale updates, $M$.}
    \label{varim}
    \end{minipage}
    \begin{minipage}[t]{0.47\textwidth}
      \centering
      \includegraphics[width=7.3cm]{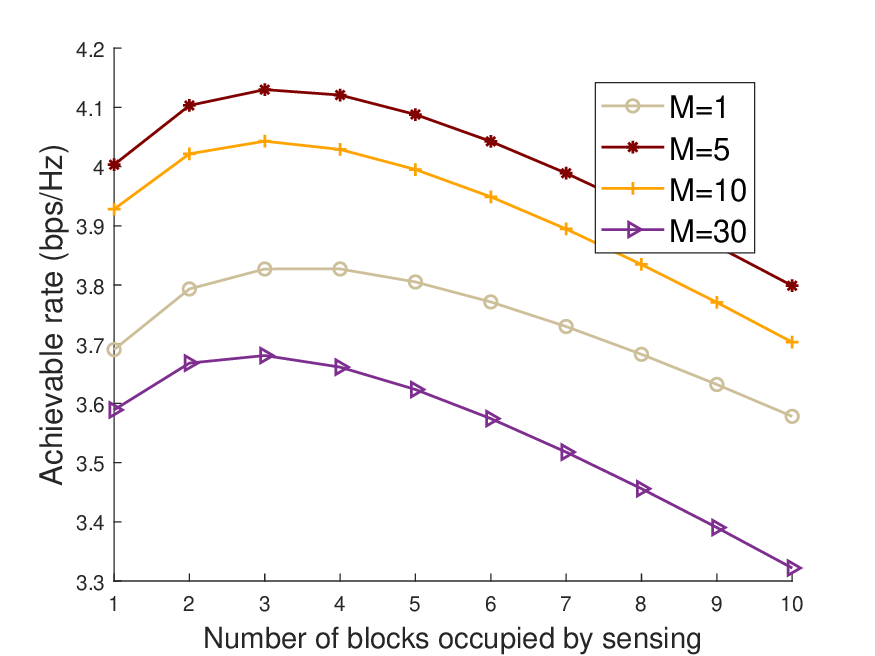}
      \caption{Achievable rate  versus the number of\\ \quad \qquad   blocks occupied by sensing, $\lfloor\frac{T_r}{T_b} \rfloor$.}
      \label{varitr}
      \end{minipage}
      \vspace{-1em}
  \end{figure*}

\subsubsection{Performance with Different Temporal Correlation Coefficient, $\rho _1$}
Fig. \ref{comrou} shows the achievable rate comparison versus the temporal correlation coefficient, $\rho _1$. It is observed that our proposed method can obtain the highest achievable rate compared with the other counterparts. Meanwhile, it also shows that, for the ``FSU'' method, the variations of $\rho _1$ do not impact the performance of the achievable rate. This makes sense because frequent small-scale updates eliminate the small-scale block fading, and the optimization design is not affected by $\rho _1$.  For all the other methods, it can be observed that the achievable rate increases with the increase of the temporal correlation coefficient. A small value of the temporal correlation coefficient corresponds to the scenario of high mobility, which means a severe small-scale fading. Therefore, the achievable rate is low with a small value of the temporal correlation coefficient.  When $\rho _1=1$, a single small-scale update suffices since there is no block fading. Under this condition, both ``RBA'' and ``SSU'' adopt the same optimization strategy as our proposed method, resulting in the same system performance. However, as  $\rho _1$ decreases, our proposed method exhibits a growing performance advantage over other counterparts.
\subsubsection{Performance with Different Small-Scale Channel Estimation Duration, $Q_e$}
$Q_e$ denotes the number of symbols in each small-scale channel estimation. Fig. \ref{compayload} shows the achievable rate versus different values of $Q_e$.  It can be seen that our proposed method outperforms other counterparts. It is noteworthy that the achievable rate of the ``FSU'' method first increases and then decreases as $Q_e$ increases, while the other methods consistently increase with $Q_e$. It reflects the system's trade-off between the sensing duration and communication time. The increase of $Q_e$ means a longer time duration for each small-scale estimation, which improves the communication efficiency due to the more accurate channel estimation while it leads to the reduction of communication time. 
When the value of $Q_e$ is in the range of 15 to 25 shown in Fig. \ref{compayload}, the benefit from the more accurate small-scale channel estimation can pay off the sacrifice of communication time, therefore, the achievable rate increases with the increase of $Q_e$. When the value of $Q_e$ further increases, the achievable rate decreases with the scheme "FSU", since both the increase of $Q_e$ and frequent small-scale updates lead to a huge sacrifice of communication time, which cannot be compensated by the benefit from the improvement in the channel estimation accuracy. Therefore, the system achievable rate decreases with the scheme “FSU”.  For other schemes, it is observed that the achievable rate keeps increasing when $Q_e$ is in the range of 25 to 45, as shown in Fig. 7. This is because other schemes can trade-off between the sensing time and communication time by reducing the update frequency.
Specifically, at $Q_e=25$, $Q_e=35$, and $Q_e=45$, the numbers of small-scale update times are 10, 9, and 7, respectively.

  \begin{figure*}[t]
  \centering
      \begin{minipage}[t]{0.47\textwidth}
        \hspace{0.5em}
        \includegraphics[width=7.7cm]{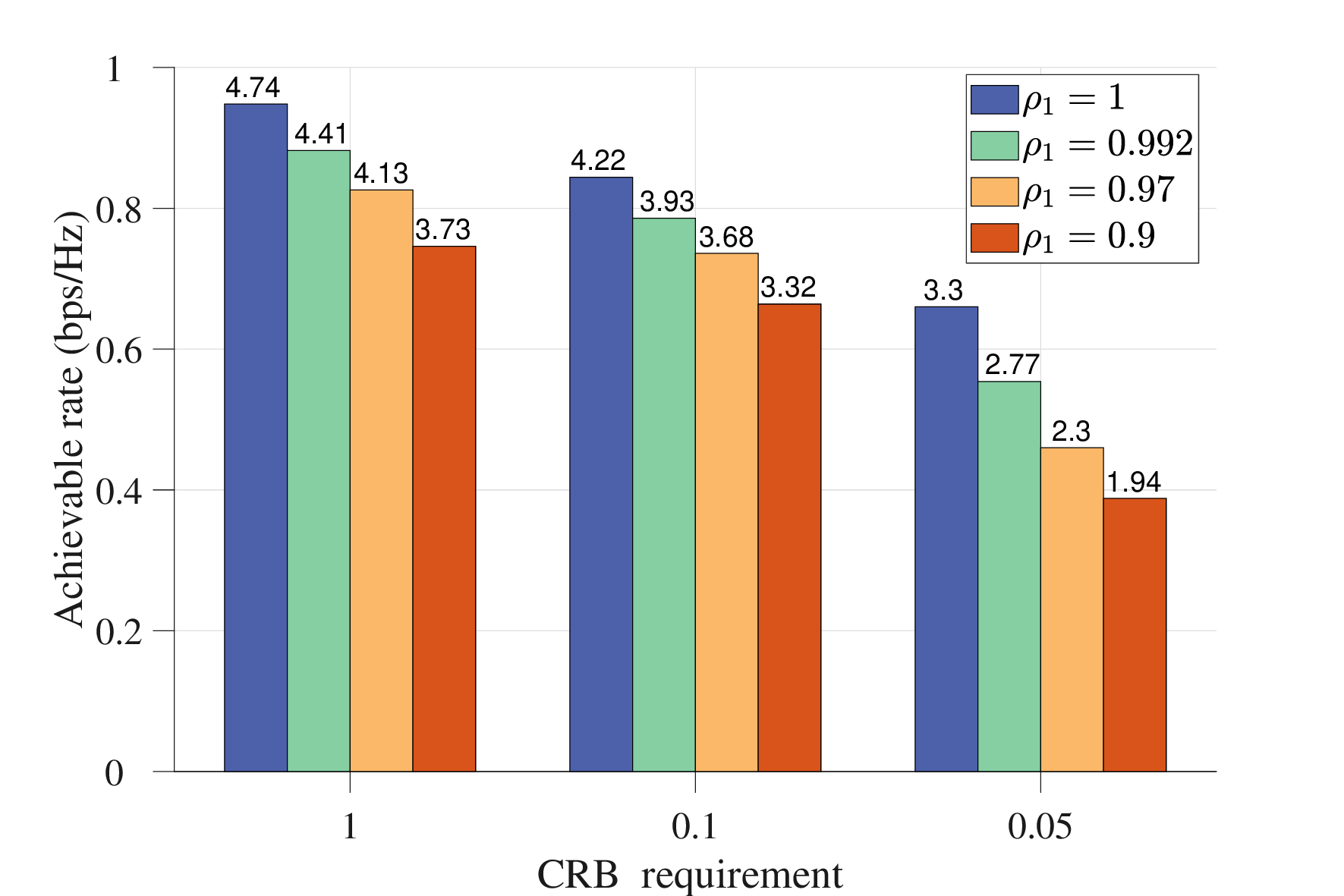}
        \caption{Achievable rate versus CRB requirement  $\varGamma$ for different temporal correlation coefficient, $\rho _1$.}
        \label{varirouthrou}
      \end{minipage} 
      \begin{minipage}[t]{0.47\textwidth}
        \hspace{1.4em}
          \includegraphics[width=7.7cm]{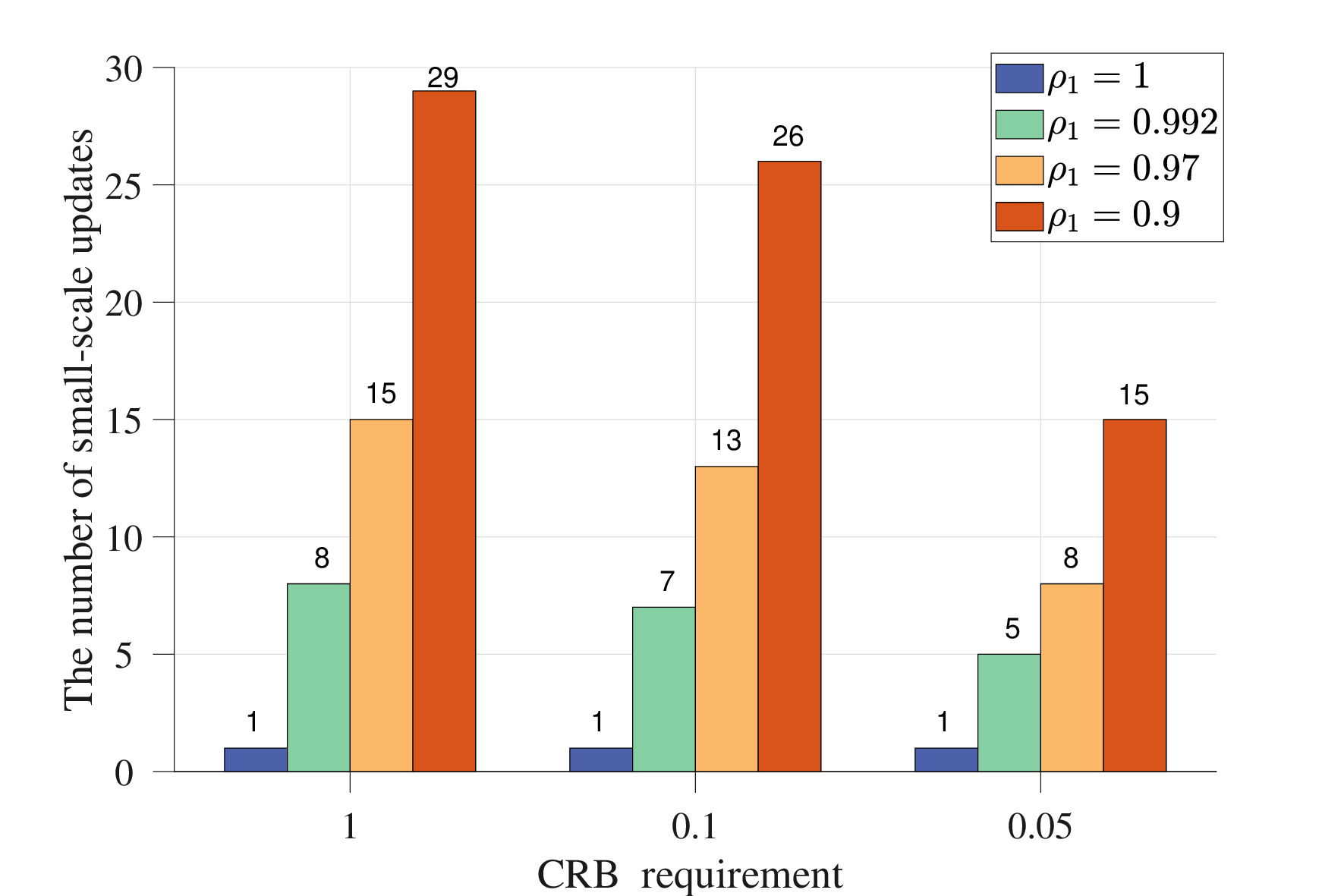}
          \caption{The number of small-scale updates versus CRB requirement  $\varGamma$ for different temporal correlation coefficient, $\rho _1$.}
          \label{variroum}
      \end{minipage}
      \vspace{-1em}
  \end{figure*}
\subsection{Impact of System Parameters}
Figs. \ref{varim} - \ref{variroum} comprehensively demonstrate the impacts of several system parameters, i.e., the number of small-scale updates, $M$; the number of blocks occupied by sensing, $\lfloor\frac{T_r}{T_b} \rfloor$; and the temporal correlation coefficient, $\rho _1$.\par
Fig. \ref{varim} depicts the achievable rate versus different numbers of small-scale updates. It is observed that the achievable rate initially increases and then decreases as the number of small-scale updates grows.  This is due to a fundamental system performance trade-off: while increasing $M$ enhances communication quality, it also reduces the time available for data transmission. Both insufficient and overly frequent small-scale updates degrade  the system performance, highlighting the importance of our proposed optimization design for $M$. Fig. \ref{varim}  further shows that the system achieves optimal performance when $\lfloor\frac{T_r}{T_b} \rfloor=3$. Additionally, the performance for $\lfloor\frac{T_r}{T_b} \rfloor=1$ is better than that for $\lfloor\frac{T_r}{T_b} \rfloor=20$ but inferior to that for $\lfloor\frac{T_r}{T_b} \rfloor=10$, indicating the necessity of  optimizing $\lfloor\frac{T_r}{T_b} \rfloor$.  Fig. \ref{varitr} shows similar trends for different numbers of blocks occupied by sensing,  i.e., $\lfloor\frac{T_r}{T_b} \rfloor$. As the system needs to balance the communication efficiency and the transmission time, the achievable rate first increases and then decreases with the growth of $\lfloor\frac{T_r}{T_b} \rfloor$. This validates the necessity of  optimizing $\lfloor\frac{T_r}{T_b} \rfloor$. Specifically, Fig. \ref{varitr}  indicates that the system achieves optimal performance when $M=5$ and  $\lfloor\frac{T_r}{T_b} \rfloor=3$, aligning with the findings in Fig. \ref{varim}.\par 

Figs. \ref{varirouthrou} and \ref{variroum} depict the achievable rate and the number of small-scale updates, respectively, versus the CRB requirement $\varGamma$, where a lower value of $\varGamma$ means a more stringent sensing requirement. It can be observed in Fig.~\ref{varirouthrou} that the achievable rate decreases with the increase of the sensing requirement (i.e., as $\varGamma$ decreases).  Achieving higher sensing accuracy requires more time resources, thereby reducing the resources available for communication and resulting in a lower achievable rate. Meanwhile, for a given CRB requirement, a smaller temporal correlation coefficient $\rho _1$ results in a lower achievable rate. This is because the small-scale fading loss increases as  $\rho _1$ decreases, i.e.,  $\rho _n=\left( \rho _1 \right) ^n$. Fig. \ref{variroum} shows that the number of small-scale updates decrease with the increase of sensing requirements. More precise sensing requirement requires longer sensing time, resulting in fewer remaining transmission time and consequently reducing the need of a large number of small-scale updates.  It is also observed that, with a given CRB requirement, the number of small-scale updates increase with $\rho _1$ decreases. A smaller value of the temporal correlation coefficient  $\rho _1$ means a more severe small-scale fading loss, which leads to a greater need for frequent small-scale updates. It can also be observed that when $\rho _1=1$, there is no block fading and only a single small-scale update is required.

\begin{figure}[htbp]
  \centerline{\includegraphics[width=8cm]{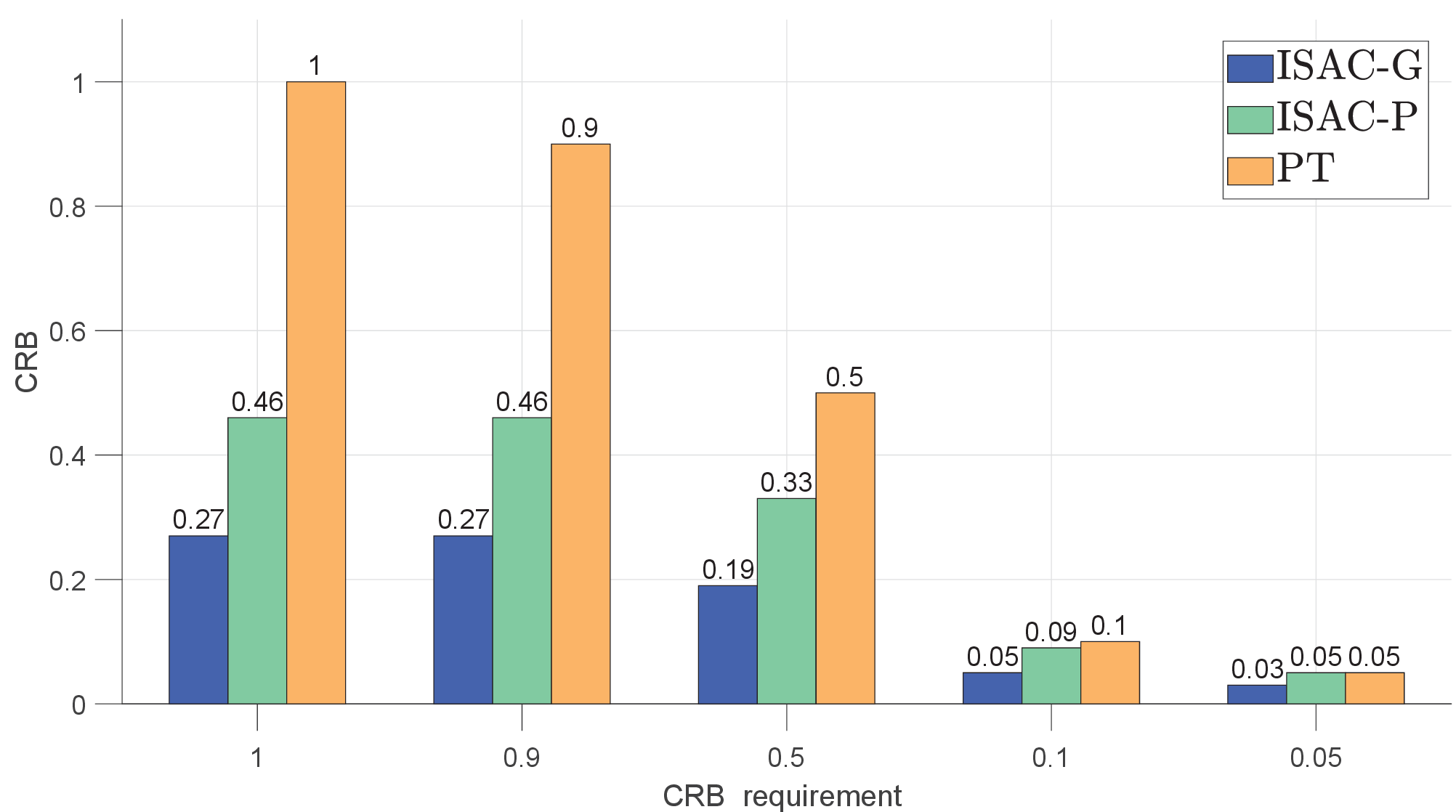}}
  \caption{CRB of different targets versus the CRB requirement $\varGamma$.}
  \label{trade34}
  \vspace{-1.5em}
\end{figure}
\subsection{Sensing-assisted Communication Efficiency}
We further demonstrate the benefits of accurate sensing in the improvement of communication efficiency. 
Fig. \ref{trade34} depicts the achieved CRB versus the CRB requirement $\varGamma$ for three users: the first one is an ISAC user away from BS with the distance of 22.69 m (denoted as ISAC-G since this user is near to BS with good channel condition); The second one is an ISAC user away from BS with the distance of 45.45 m (denoted as ISAC-P since this user is far away from BS with poor channel condition compared with ISAC-G); The third one is the pure target (denoted as PT).  
It can be observed that both the pure target and the ISAC users satisfy the sensing requirements.  In addition, for ISAC users, the achieved CRB values are lower than the CRB requirement $\varGamma$.
That is, the actual CRBs of ISAC users are better (i.e., lower) than the CRB requirement. This is because more accurate sensing can be exploited to improve communication efficiency. More accurate sensing can enhance communication channel estimation and consequently improve communication efficiency. The objective of ISAC users is to maximize their communication efficiency while satisfying the sensing requirement. Thus, the optimal CRBs of both ISAC users are lower than the CRB requirement. Fig. \ref{trade34} also shows that the optimal CRB values for ISAC-G are lower than those for ISAC-P.  A lower CRB corresponds to higher sensing accuracy. Since ISAC-G has better channel condition than ISAC-P, prioritizing more accurate sensing for ISAC-G yields greater improvements in the overall system communication efficiency. Finally, through the optimization of radar transmit beamformers and sensing duration, the optimal values of CRB are different among ISAC users.  Our proposed method can take the best advantage of accurate sensing to maximize the sensing-assisted communication.

\section{Conclusion}\label{Conclusion}
In this paper, we have investigated the optimal design of dual-scale channel estimation for sensing-assisted communication systems. Considering the impact of sensing detection errors on large-scale fading and the channel aging effect on small-scale fading, we have explored the joint optimization of sensing duration, small-scale update timing and frequency, radar transmit beamformers, and communication power via a structured frame design. Then, we have formulated an average achievable rate maximization problem and proposed a corresponding segmented golden search-based algorithm to solve it efficiently.  Simulation results have demonstrated that our optimal design can achieve the improvement of 10.7\%, 14.4\%, 24.3\%, and 152.4\%  in comparison with the schemes of equal power allocation, random block allocation, frequent small-scale update, and single small-scale update, respectively. To improve the system performance, the cooperative interplay between sensing and communication has also been highlighted. For future work, we plan to explore dynamic time management via reinforcement learning  in dual-scale channel estimation.
  	\begin{appendices}    
\section{Proof of Proposition 1} 
    With the MRT beamforming vector $\mathbf{f}_{k,n}=\frac{1}{\sqrt{\mathrm{tr}\left( \boldsymbol{C}_{h,k} \right)}}\tilde{\boldsymbol{h}}_{k,1}$, terms in (\ref{82632}) can be  summarized as
    \begin{itemize}
      \item [1)]  Compute $\mathbb{E} \left[ \boldsymbol{h}_{k,n}^{H}\mathbf{f}_{k,n}\right]$: It is general to assume that $\bar{\boldsymbol{e}}_{k, n}$ and $\tilde{\boldsymbol{h}}_{k,1}$ are statistically independent\cite{zheng2021impact,li2017channel}. Therefore, we have  
      \begin{equation}
      \begin{split}
        \mathbb{E} & \left[ \boldsymbol{h}_{k,n}^{H}\mathbf{f}_{k,n} \right]\\
          &=\mathbb{E} \left[ \left( \rho _{k,n}\tilde{\boldsymbol{h}}_{k,1}^H+\bar{\boldsymbol{e}}_{k, n}^H \right) \frac{\tilde{\boldsymbol{h}}_{k,1}}{\sqrt{\mathrm{tr}\left( \boldsymbol{C}_{h,k} \right)}} \right]\\
          &= \mathbb{E} \left[ \frac{\rho _{k,n}\tilde{\boldsymbol{h}}_{k,1}^H\tilde{\boldsymbol{h}}_{k,1}}{\sqrt{\mathrm{tr}\left( \boldsymbol{C}_{h,k} \right)}} \right] =\rho _{k,n}\sqrt{\mathrm{tr}\left( \boldsymbol{C}_{h,k} \right)}.
      \end{split}
      \end{equation}
      \item [2)]  Compute $\mathbb{E} \left[ \left| \boldsymbol{h}_{k,n}^{H}\mathbf{f}_{i,n} \right|^{2} \right]$, for $i\ne k$:
      \vspace{-1em}
      \begin{equation}
      \begin{split} 
        \mathbb{E} \left[ \left| \boldsymbol{h}_{k,n}^{H}\mathbf{f}_{i,n} \right|^{2} \right]& =\frac{\mathbb{E} \left[ \boldsymbol{h}_{k,n}^{H}\tilde{\boldsymbol{h}}_{i,1}\tilde{\boldsymbol{h}}_{i,1}^{H}\boldsymbol{h}_{k,n} \right]}{\mathrm{tr}\left( \boldsymbol{C}_{h,i} \right)}
        \\
       &=\! \frac{\mathrm{tr}\left( \mathbb{E} \left[ \boldsymbol{h}_{k,n}\boldsymbol{h}_{k,n}^{H} \right]\! \mathbb{E} \left[ \tilde{\boldsymbol{h}}_{i,1}\tilde{\boldsymbol{h}}_{i,1}^{H} \right] \right)}{\mathrm{tr}\left( \boldsymbol{C}_{h,i} \right)}
       \\
       &=\frac{\beta _k\mathrm{tr}\left( \boldsymbol{R}_{k}^{S}\boldsymbol{C}_{h,i} \right)}{\mathrm{tr}\left( \boldsymbol{C}_{h,i} \right)}.
      \end{split}
      \end{equation}
      \vspace{-1em}
      \item [3)]  Compute $\mathbb{E} \left[ \left| \boldsymbol{h}_{k,n}^{H}\mathbf{f}_{k,n}-\mathbb{E} \left[ \boldsymbol{h}_{k,n}^{H}\mathbf{f}_{k,n} \right] \right|^{2} \right]$: 
      \begin{equation}
        \begin{split} 
      \mathbb{E}& \left[ \left| \boldsymbol{h}_{k,n}^{H}\mathbf{f}_{k,n}-\mathbb{E} \left[ \boldsymbol{h}_{k,n}^{H}\mathbf{f}_{k,n} \right] \right|^{2} \right]\\
      =&\mathbb{E} \left[ \left| \frac{\rho _{k,n}\tilde{\boldsymbol{h}}_{k,1}^{H}\tilde{\boldsymbol{h}}_{k,1}}{\sqrt{\mathrm{tr}\left( \boldsymbol{C}_{h,k} \right)}} \right|^2 \right] +\mathbb{E} \left[ \left| \frac{\tilde{\boldsymbol{h}}_{k,1}^{H}\bar{\boldsymbol{e}}_{k, n}}{\sqrt{\mathrm{tr}\left( \boldsymbol{C}_{h,k} \right)}} \right|^2 \right] \\
      &- \rho _{k,n}^{2}\mathrm{tr}\left( \boldsymbol{C}_{h,k} \right)\\
      =&\frac{\rho _{k,n}^{2}\mathbb{E} \left[ \tilde{\boldsymbol{h}}_{k,1}^{H}\tilde{\boldsymbol{h}}_{k,1}\tilde{\boldsymbol{h}}_{k,1}^{H}\tilde{\boldsymbol{h}}_{k,1} \right]}{\mathrm{tr}\left( \boldsymbol{C}_{h,k} \right)}+\frac{\mathrm{tr}\left( \boldsymbol{C}_{h,k}\bar{\boldsymbol{C}}_{e,k}\right)}{\mathrm{tr}\left( \boldsymbol{C}_{h,k} \right)}\\
      &-\rho _{k,n}^{2}\mathrm{tr}\left( \boldsymbol{C}_{h,k} \right),
      \end{split}
      \end{equation}
    where $\bar{\boldsymbol{C}}_{e,k}=\beta _k \boldsymbol{R}_{k}^{S}-\rho _{k,n}^2\boldsymbol{C}_{h,k}$
     Then, if the MMSE estimation is deployed, we have
      \begin{equation}
        \begin{split} 
          &\mathbb{E} \left[ \tilde{\boldsymbol{h}}_{k,1}^{H}\tilde{\boldsymbol{h}}_{k,1}\tilde{\boldsymbol{h}}_{k,1}^{H}\tilde{\boldsymbol{h}}_{k,1} \right] 
          \\
          &=\mathbb{E} \left[ \small{\left| \frac{1}{T_{e}^{2}P_e}\mathbf{y}_{e,k}^{H}\dot{\boldsymbol{R}}_{k}^{-1}\hat{\boldsymbol{R}}_k\hat{\boldsymbol{R}}_k\dot{\boldsymbol{R}}_{k}^{-1}\mathbf{y}_{e,k} \right|^2} \right]           \\
          &=\left| \mathrm{tr}\left(\dot{\boldsymbol{R}}_{k}^{-1} \hat{\boldsymbol{R}}_k\hat{\boldsymbol{R}}_k\dot{\boldsymbol{R}}_{k}^{-1}\dot{\boldsymbol{R}}_k \right) \right|^2\\
          &+ \! \mathrm{tr}\left( \dot{\boldsymbol{R}}_{k}^{-1}\hat{\boldsymbol{R}}_k\hat{\boldsymbol{R}}_k\dot{\boldsymbol{R}}_{k}^{-1}\dot{\boldsymbol{R}}_k\left( \dot{\boldsymbol{R}}_{k}^{-1}\hat{\boldsymbol{R}}_k\hat{\boldsymbol{R}}_k\dot{\boldsymbol{R}}_{k}^{-1} \right)^{H} \!\! \dot{\boldsymbol{R}}_k \right)\\
          &=\left| \mathrm{tr}\left( \boldsymbol{C}_{h,k} \right) \right|^{2}+\mathrm{tr}\left( \boldsymbol{C}_{h,k}\boldsymbol{C}_{h,k} \right),  
      \end{split}
      \end{equation}
      where $\dot{\boldsymbol{R}}_k=\left( \boldsymbol{\hat{R}}_k+\frac{1}{\gamma _e}\mathbf{I}_{\text{L}_{\text{t}}} \right)$.  We then conclude that
      \begin{equation}
      \begin{split} 
      \mathbb{E} &\left[ \left| \boldsymbol{h}_{k,n}^{H}\mathbf{f}_{k,n}-\mathbb{E} \left[ \boldsymbol{h}_{k,n}^{H}\mathbf{f}_{k,n} \right] \right|^{2} \right]\\
      & =\frac{\rho _{k,n}^{2}\mathrm{tr}\left( \boldsymbol{C}_{h,k}\boldsymbol{C}_{h,k} \right) +\mathrm{tr}\left( \boldsymbol{C}_{h,k}\bar{\boldsymbol{C}}_{e,k}
      \right)}{\mathrm{tr}\left( \boldsymbol{C}_{h,k} \right)}\\
      &=\frac{\beta _k\mathrm{tr}\left( \boldsymbol{C}_{h,k}\boldsymbol{R}_{k}^{S} \right)}{\mathrm{tr}\left( \boldsymbol{C}_{h,k} \right)}.
    \end{split}
    \end{equation} 
    \end{itemize}
    \quad Finally,  the  $\gamma _{k,n}$  in (\ref{82632}) can be represented as
      \begin{equation}
        \begin{split} 
          \gamma _{k,n}\!=\!\frac{p_k\rho _{k,n}^{2}\mathrm{tr}\left( \boldsymbol{C}_{h,k} \right)}{\sum_{i=1}^K\frac{\beta _kp_i\mathrm{tr}\left( \boldsymbol{R}_{k}^{S}\boldsymbol{C}_{h,i} \right)}{\mathrm{tr}\left( \boldsymbol{C}_{h,i} \right)}\!+\!\sigma _{c}^{2}},\forall k.
        \end{split}
        \end{equation}
    \label{apdxb}

\section{Proof of Proposition 2}    
\label{apdxc}
Based on (\ref{92637}), Lagrangian function of $\mathbf{P}2$ is given by 
\begin{equation}
  \begin{split} 
\mathcal{L} \left( N_m,\lambda _m,\mu\right) =&- {\frac{1}{N}}\sum_{m=1}^M{\!}\sum_{n=1}^{N_m}{\varPhi _n}+\sum_{m=1}^M{\!}\lambda _m\left( 1-N_m \right)\\
& +\mu\left( \sum_{m=1}^M{N_m}+\lfloor \frac{T_r}{T_b} \rfloor -N \right). 
\end{split}
\end{equation}
To satisfy the stability condition, we can get
\begin{equation}
  \begin{split} 
\small{\frac{\partial \mathcal{L} \left( N_m,\lambda _m,\mu\right)}{\partial N_m}}=-{\frac{\varPhi _{N_m}}{N}}-\lambda _m+\mu=0.
\end{split}
\end{equation}
Whether the optimal solution is on the boundary or not, the complementary slackness condition, $\lambda _m\left( 1-N_m \right) =0$, helps us to conclude that
\begin{equation}
  \begin{split} 
    \left( \mu-{\frac{\varPhi _{N_m}}{N}} \right) \left( 1-N_m \right) =0.
\end{split}
\end{equation}
Then, based on $M\leqslant N$ and $\varPhi _1>\varPhi _2,\cdots,>\varPhi _{N_m}$, the rapid update of $N_m=1$ will inevitably cause a decrease in communication achievable rate. Thus, we have $N_m=\frac{N_t}{M},\forall m$. \par
However, $N_m$ is a discrete variable and we have $\varPhi _1>\varPhi _2,\cdots,>\varPhi _{N_m}$. This signifies that, with each update comprising $ \lfloor \frac{N_t}{M} \rfloor$ blocks, $\mathrm{mod}\left( \frac{N_t}{M} \right)$ redundant blocks need to be spread evenly in the $M$ updates. Thus, there are $\mathrm{mod}\left( \frac{N_t }{M} \right) $ updates that have $\frac{N_t}{M}+1$ block and we can conclude that
\begin{equation}
  \begin{split} 
     N_m\in \left\{ \lfloor \frac{N_t}{M} \right. \rfloor ,\;\left. \lfloor \frac{N_t}{M}\rfloor +1 \right\}, \forall m.
  \end{split}
  \end{equation}
\section{Proof of Proposition 3}    
\label{apdxd}
Assuming the number of blocks occupied by $T_r$ is fixed,  $\lfloor \frac{N_t}{M} \rfloor$ will be a constant value and $\mathrm{mod}\left( T_r,T_b \right)$ can be represented as $\hat{T}_r$, which is continuous. Then, only $\mathrm{SE}_{k,n}$ is correlated with  $\hat{T}_r$ in the first two terms of (\ref{101160}). Based on (\ref{101039}), the derivative of  $\mathrm{SE}_{k,n}$ is given by\par  
 \begin{equation}
  \begin{split} 
    \label{102942}
    \frac{\partial \mathrm{SE}_{k,n}}{\partial \hat{T}_r}
=\frac{\mathrm{ln}2\kappa _{k,n}}{1+\kappa _{k,n}\mathrm{tr}\left( \boldsymbol{C}_{h,k} \right)}\frac{\partial \mathrm{tr}\left( \boldsymbol{C}_{h,k} \right)}{\partial \hat{T}_r},\forall k,
\end{split}
\end{equation}
where $\kappa _k=\frac{p_k\rho _{k,n}^{2}}{\beta _kP_m\mathrm{tr}\left( \boldsymbol{R}_{k}^{S} \right)\!+\!\sigma _{c}^{2}}
$ is derived from (\ref{101039}). The channel estimation quality is reflected in $\boldsymbol{C}_{h,k}$, for which we have the simpler form as follows
\begin{equation}
  \begin{split} 
    \label{101040}
\mathrm{tr}\left( \boldsymbol{C}_{h,k} \right) =\mathrm{tr}\left( \hat{\boldsymbol{R}}_k \right) +\frac{1}{\gamma _{e}^{2}}\mathrm{tr} \left( \hat{\boldsymbol{R}}_k+\frac{1}{\gamma _e}\mathbf{I}_{{L}_{{t}}} \right) ^{-1}\!\!\!-\frac{{L}_{{t}}}{\gamma_e},\forall k.
\end{split}
\end{equation} With the identity of the Neuman series for matrices\cite{andre2020joint}, the following approximation is given 
\begin{equation}
  \begin{split} 
\frac{1}{\gamma _{e}^{2}}\mathrm{tr}\left( \hat{\boldsymbol{R}}_k+\frac{1}{\gamma _e}\mathbf{I}_{{L}_{{t}}} \right) ^{-1}\cong \mathrm{tr}\left( \frac{\mathbf{I}_{{L}_{{t}}}}{\gamma _e}-\hat{\boldsymbol{R}}_k+\gamma _e\hat{\boldsymbol{R}}_k\hat{\boldsymbol{R}}_k \right),\forall k. 
\end{split}
\end{equation}
Then, we can conclude that 
\begin{equation}
  \begin{split} 
    \label{101163}
\mathrm{tr}\left( \boldsymbol{C}_{h,k} \right)&\cong  \gamma _e\mathrm{tr}\left(\hat{\boldsymbol{R}}_k\hat{\boldsymbol{R}}_k\right)\\
&=\gamma _e\mathrm{CRB}^2\left( \theta _k \right) \mathrm{tr}\!\left(\! \tilde{\boldsymbol{R}}_{r,k}\tilde{\boldsymbol{R}}_{r,k} \right)\!+\!\gamma _e\beta _{k}^{2}\mathrm{tr}\!\left( \boldsymbol{R}_{k}^{S}\boldsymbol{R}_{k}^{S} \right) \\ &-2\gamma _e\mathrm{CRB}\left( \theta _k \right) \beta _k\mathrm{tr}\!\left( \tilde{\boldsymbol{R}}_{r,k}\boldsymbol{R}_{k}^{S} \right), \forall k,
\end{split}
\end{equation}
where $\tilde{\boldsymbol{R}}_{r,k}=\boldsymbol{f}\prime\left( \theta _k \right) \boldsymbol{f}\prime\left( \theta _k \right) ^H$ comes from (\ref{82920b}), and $\mathrm{tr}\!\left( \tilde{\boldsymbol{R}}_{r,k}\boldsymbol{R}_{k}^{S} \right)$ is a real value since $\tilde{\boldsymbol{R}}_{r,k}$ and $\boldsymbol{R}_{k}^{S}$ are Hermitian matrix. Since we assume $\hat{\boldsymbol{R}}_k=\beta _k\boldsymbol{R}_{k}^{S}-\mathrm{CRB}\left( \theta _k \right) \tilde{\boldsymbol{R}}_{r,k}\succeq 0$, the first-order derivative of (\ref{101163}) on  $\mathrm{CRB}\left( \theta _k \right)$ is given by
\begin{equation}
\begin{split} 
  &\mathcal{U} \left( \mathrm{CRB}\left( \theta _k \right) \right)\triangleq \\ 
  &\ 2\gamma _e\mathrm{CRB}\left( \theta _k \right) \mathrm{tr}\!\left( \tilde{\boldsymbol{R}}_{r,k}\tilde{\boldsymbol{R}}_{r,k} \!\right)
   \!-\!2\gamma _e\beta _k\mathrm{tr}\left( \boldsymbol{R}_{k}^{S}\tilde{\boldsymbol{R}}_{r,k} \right)\!\!<\! 0,\forall k.  
\end{split}
\end{equation}
Therefore, the maximization of $\mathrm{tr}\left( \boldsymbol{C}_{h,k} \right)$ is equivalent to minimizing $\mathrm{CRB}\left( \theta _k \right)$. Thus, the lower bound $\mathcal{U} _{lb,k}^{\left( t \right)}$ of (\ref{101163}) in $t$-th iteration can be obtained by applying the first-order Taylor approximation 
\begin{equation}
\begin{split} 
    \label{112572}
    \mathrm{tr}\left( \boldsymbol{C}_{h,k} \right) \!\geqslant & \ \mathcal{U} \left( \mathrm{CRB}^{\left( t \right)}\left( \theta _k \right) \right) \mathrm{CRB}\left( \theta _k \right) +\mathrm{tr}\!\left( \boldsymbol{C}_{h,k}^{\left( t \right)} \right)\!\!\\& -\mathcal{U} \!\left(\! \mathrm{CRB}^{\left( t \right)}\!\left( \theta _k \right) \!\right) \mathrm{CRB}^{\left( t \right)}\!\left( \theta _k \right)\triangleq \mathcal{U} _{lb,k}^{\left( t \right)},\forall k.
\end{split}
\end{equation}
Accordingly, we can conclude
\begin{equation}
  \begin{split}  
\frac{\partial\mathcal{U} _{lb,k}^{\left( t \right)}}{\partial \hat{T}_r}\!\!\propto\frac{-\mathcal{U} \left( \mathrm{CRB}^{\left( t \right)}\left( \theta _k \right) \right)}{\hat{T}_{r}^{2}}\geqslant 0,\forall k.
\end{split}
\end{equation}
Since $\mathrm{tr}\left( \boldsymbol{C}_{h,k} \right)$ increases with $\hat{T}_r$, we conclude that the derivative form in (\ref{102942}) is monotonically decreasing with $\hat{T}_r$ after approximation. Thus, $\mathrm{SE}_{k,n}$ is concave with respect to  $\hat{T}_r$. For the second row of (\ref{101160}), combined with $\mathrm{SE}_{k,1}$ in the first row, it can be transformed as $\frac{\left( MT_b-MT_e-\hat{T}_r \right)}{NT_b} \sum_{k=1}^K{\mathrm{SE}_{k,1}}$ and its derivative is given by
\begin{equation}
  \begin{split} 
    \label{101464}
-\frac{1}{NT_b} \sum_{k=1}^K{\mathrm{SE}_{k,1}}+\frac{\left( MT_b-MT_e-\hat{T}_r \right)}{NT_b} \sum_{k=1}^K \frac{\partial {\mathrm{SE}_{k,1}}}{\partial \hat{T}_r}.
\end{split}
\end{equation}
Since $\mathrm{SE}_{k,1}$ increases with $\hat{T}_r$, the expression in (\ref{101464}) is obviously decreasing with  $\hat{T}_r$ increases, thus it is concave on $\hat{T}_r$.
\end{appendices} 

\bibliographystyle{IEEEtran}
\bibliography{document.bib}
\end{document}